\documentclass[12pt,preprint]{aastex}

\def\aa#1#2#3#4#5{\bibitem[#1]{#2}#3, {A\&A}, {#4}, #5}
\def\aasup#1#2#3#4#5{\bibitem[#1]{#2}#3, {A\&AS}, {#4}, #5}
\def\aj#1#2#3#4#5{\bibitem[#1]{#2}#3, {AJ}, {#4}, #5}
\def\apj#1#2#3#4#5{\bibitem[#1]{#2}#3, {Ap. J.}, {#4}, #5}

\def\apss#1#2#3#4#5{\bibitem[#1]{#2}#3, {Ap\&SS}, {#4}, #5}
\def\apjsup#1#2#3#4#5{\bibitem[#1]{#2}#3, ApJS, #4, #5}

\def\mnras#1#2#3#4#5{\bibitem[#1]{#2}#3, {M.N.R.A.S.}, {#4}, #5}
\def\nature#1#2#3#4#5{\bibitem[#1]{#2}#3, {Nature}, {#4}, #5}

\def\pasp#1#2#3#4#5{\bibitem[#1]{#2}#3, {PASP}, {#4}, #5}

\def\spie#1#2#3#4#5{\bibitem[#1]{#2}#3, {Proc. SPIE}, {#4}, #5}

\begin{document}

\title{Near-Infrared Interferometric Measurements of Herbig Ae/Be Stars}

\author{J.A. Eisner\altaffilmark{1}, B.F. Lane\altaffilmark{2}, 
R.L. Akeson\altaffilmark{3}, L.A. Hillenbrand\altaffilmark{1}, 
\& A.I. Sargent\altaffilmark{1}}
\email{jae@astro.caltech.edu}
\altaffiltext{1}{California Institute of Technology,
Department of Astronomy MC 105-24,
Pasadena, CA 91125}
\altaffiltext{2}{California Institute of Technology,
Department of Geological and Planetary Sciences MC 150-21,
Pasadena, CA 91125}
\altaffiltext{3}{California Institute of Technology, 
Michelson Science Center MC 100-22,
Pasadena, CA 91125}

\keywords{stars:pre-main sequence---stars:individual(AB Aur,VV Ser,V1685 Cyg,
AS 442,MWC 1080)---techniques:high angular resolution}

\slugcomment{Accepted for publication in the Astrophysical Journal}

\begin{abstract}
We have observed the Herbig Ae/Be sources AB Aur, VV Ser, V1685 Cyg
(BD+40$^{\circ}$4124), AS 442,
and MWC 1080 
with the Palomar Testbed Interferometer, obtaining the
longest baseline near-IR interferometric observations of this
class of objects. All of the sources are resolved at 2.2 $\mu$m with
angular size scales generally $\la 5$ mas,
consistent with the only previous near-IR interferometric measurements
of Herbig Ae/Be stars by Millan-Gabet and collaborators.
We determine the angular size scales and orientations predicted
by uniform disk, Gaussian, ring, and accretion disk models.
Although it is difficult to distinguish different radial distributions,
we are able to place
firm constraints on the inclinations of these models, and our measurements
are the first that show evidence for significantly inclined morphologies.
In addition, the derived angular sizes for the
early type Herbig Be stars in our sample, V1685 Cyg and MWC 1080, 
agree reasonably well with those predicted by the face-on accretion disk models
used by Hillenbrand and collaborators to explain observed spectral
energy distributions.  In contrast, our data for the later-type sources
AB Aur, VV Ser, and AS 442 are somewhat inconsistent with these models, 
and may be explained better through the puffed-up inner disk models 
of Dullemond and collaborators.  
\end{abstract}

\section{Introduction \label{sec:intro}}
Herbig Ae/Be \citep[HAEBE;][]{HERBIG60}
stars are intermediate-mass (2--10 M$_{\odot}$) young 
stellar objects
that show broad emission lines, rapid variability, and
excess infrared and millimeter-wavelength emission. 
These properties are consistent with 
the presence of hot and cold circumstellar dust and gas.  
While there is still some debate about the morphology of this circumstellar
material, most evidence supports the hypothesis that in many cases
the dust and gas lies in a massive ($\sim 0.01$ M$_{\odot}$) 
circumstellar disk
(Natta, Grinin, \& Mannings 2000; Hillenbrand et al. 1992, hereafter HSVK).

The strongest evidence for circumstellar disks around HAEBE stars comes
from direct imaging with millimeter interferometry.  Flattened structures 
around several sources have been resolved 
on $\sim 100$ AU scales \citep{MS97,MS00,PDK03}, and detailed
kinematic modeling of one source, MWC 480, shows that the observations
are fit well by a rotating Keplerian disk \citep{MKS97}. 
For a spherical distribution, these and other observations
\citep[e.g.,][]{MANNINGS94} imply extinctions at visible and infra-red
wavelengths much higher than actually observed.
In addition, in recent H$\alpha$ spectropolarimetric observations of HAEBE 
sources (which trace dust on scales of tens of stellar radii), Vink et al.
(2003) find signatures of flattened circumstellar structures around
83\% of their sample, and evidence for rotation around 9 HAe stars.
Furthermore, the
forbidden emission lines that arise in winds and outflows around
HAEBE sources typically show
blue-shifted emission but lack redshifted emission, which suggests
that the redshifted component of the outflow is occluded by a 
circumstellar disk.  The broad linewidths of low-velocity
features are consistent with this emission arising in rotating circumstellar
disk winds \citep{CR97}.

The distribution of circumstellar material around
HAEBEs can also be inferred from 
modeling of spectral energy distributions (SEDs).
Three distinct morphologies were identified in this way by
HSVK, who classified
observed HAEBE sources into three groups, I, II, and III.
All sources in our observed sample fall into Group I, which has
SEDs of the form $\lambda F_{\lambda} \propto \lambda^{4/3}$. These can be
modeled well by flat, irradiated, accretion disks with
inner holes on the order of $\sim 10$ stellar radii.
Recent SED modeling of a sample of fourteen isolated HAEBE stars with the
characteristics of Group I sources is 
consistent with emission from a passive reprocessing disk
\citep{MEEUS+01}.  Moreover, Meeus et al. (2001) 
\citep[and other investigators, e.g.,][]{NATTA+01}
attribute this emission to
the outer part of a flared circumstellar disk \citep[e.g.,][]{CG97},
while previous authors attributed blackbody components
observed in SEDs of HAEBE sources to tenuous envelopes
\citep{HARTMANN+93,MIROS+99,NATTA+93}.


Size scales and orientations of disks around HAEBE stars can only
be determined directly
through high angular resolution imaging.
The spatial and velocity structure of cooler outer
HAEBE disks on $\sim 100$ AU scales
has been mapped with millimeter-wave interferometers 
(as discussed above).
To probe the warmer inner regions of the disk ($\sim 1$ AU scales),
measurements with near-IR interferometers are necessary.
The only
near-IR interferometric observations of HAEBE sources to date, conducted 
with the IOTA interferometer (Millan-Gabet et al. 1999; Millan-Gabet,
Schloerb, \& Traub 2001, hereafter MST),
led to sizes and orientations of
sources largely inconsistent with values estimated using other
techniques.  A geometrically flat disk 
may be too simplistic to accommodate all the observations, and 
puffed up inner disk walls (Dullemond, Dominik, \& Natta 2001; hereafter DDN) 
or flared outer disks  \citep[e.g.,][]{CG97} may need to be included in 
the models.  However, the limited number of HAEBE sources observed
with near-IR interferometers and the sparse $u-v$ coverage of these
observations \citep{M-G+01} make it difficult to draw unambiguous conclusions
about the structure of the circumstellar material.

We have begun a program with the Palomar Testbed Interferometer (PTI)
to observe HAEBE stars. By increasing
the sample size and improving $u-v$ coverage, we aim 
to understand better the structure
of the circumstellar emission on $\sim 1$ AU scales. 
In this paper, we present results for five sources, AB Aur,
VV Ser, V1685 Cyg (BD+40$^{\circ}$4124), AS 442, and MWC 1080.  
We note that neither our list of HAEBEs, nor that of MST, 
represents an unbiased sample, but rather, is limited to those
stars that are bright enough ($K \la 6.5$) to be 
successfully observed. 
We model the structure
of circumstellar dust around HAEBE stars
using these PTI data, together with IOTA measurements
where available.  Specifically, we compare various models---Gaussians,
uniform disks, uniform rings, and accretion disks with inner holes---to 
the visibility data to determine approximate
size scales and orientations of the circumstellar emission.  

In \S \ref{sec:obs}, we describe the PTI observations.
In \S \ref{sec:res}, we fit the observed data to several different
models for the circumstellar dust distribution and derive angular sizes
and orientations. Implications of the modeling and 
comparisons with previous observations are discussed in 
\S \ref{sec:disc}.

\section{Observations and Calibration \label{sec:obs}}
The Palomar Testbed Interferometer (PTI) 
is a long-baseline near-IR Michelson interferometer 
located on Palomar Mountain near San Diego, CA \citep{COLAVITA+99}.
PTI combines starlight from two 40-cm aperture telescopes using
a Michelson beam combiner, and records the resulting 
fringe visibilities.
These fringe visibilities are related to the
source brightness distribution via the van Cittert-Zernike theorem,
which states that the visibility distribution in $u-v$ space 
and the brightness distribution on the sky are Fourier transform 
pairs \citep{BW99}.

We observed five HAEBE sources, AB Aur, VV Ser, V1685 Cyg 
(BD+40$^{\circ}$4124), AS 442, and MWC 1080,  with 
PTI between May and October of 2002.  Properties of the sample
are included in Table \ref{tab:sources}.
We obtained K-band (2.2 $\mu$m) measurements
on an 85-m North-West (NW)  baseline for all five objects,
and on a 110-m North-South (NS) baseline for three.
The NW baseline is oriented $109^{\circ}$ west of north and has a fringe 
spacing of $\sim 5$ mas, and the NS baseline is $160^{\circ}$ west of
north and has a fringe spacing of $\sim 4$ mas.
A summary of the observations is given in Table \ref{tab:obs}.

PTI measures fringes in two channels, corresponding to the
two outputs from the beam combiner.  One output is spatially filtered
with an optical fiber
and dispersed onto five ``spectral'' pixels, while the other output
is focused onto a single ``white-light'' pixel (without spatial filtering).
The white-light pixel is used principally for fringe-tracking: the
fringe phase is measured and then used to
control the delay line system to track atmospheric fringe motion
(and thus maintain zero optical path difference between the 
two interfering beams).  The spectral pixels are generally used to
make accurate measurements of the squared visibility amplitudes ($V^2$) of
observed sources.
We sample the data at either 20 or 50 milliseconds in order to make
measurements on a timescale shorter than the atmospheric coherence time.
A ``scan'', which is the unit of data we will use in the analysis below,
consists of 130 seconds of data, divided into five 
equal time blocks.
The $V^2$ is calculated for each of these blocks using an incoherent
average of the constituent 20 or 50-ms measurements
from a synthetic wide-band channel formed from the five spectral 
pixels \citep{COLAVITA99}.   $V^2$ for the entire
scan is given by the mean of these five estimates, and the statistical
uncertainty is given by the standard deviation from the mean value.

We calibrate the measured $V^2$ for the observed HAEBE 
sources by comparing them to visibilities measured for
calibrator sources of known angular sizes, 
for which we can easily calculate the
expected $V^2$ for an ideal system.  
The visibilities are normalized such that $V^2=1$ for a point source
observed with an ideal system.  We calculate
the expected $V^2$ by assuming that the calibrators are uniform stellar
disks.  Making use of the van Cittert-Zernike theorem,
the squared visibilities for these sources are given by
\begin{equation}
V^2_{\rm calc} = \left( 2 \frac{J_1(\pi \theta r_{\rm uv})}
{\pi \theta r_{\rm uv}} 
\right)^2. 
\label{eq:ud}
\end{equation}
Here, $J_1$ is the first-order Bessel function.
$\theta$ is the angular diameter of the star, and $r_{\rm uv}$ is the
``uv radius'', defined by
\begin{equation}
r_{\rm uv} = \sqrt{u^2+v^2} = \frac{\vec{B} \cdot \vec{s}}{\lambda},
\label{eq:ruv}
\end{equation}
where $\vec{B}$ is the baseline vector, $\vec{s}$ is a unit vector 
pointing from the center of the baseline towards the source, 
and $\lambda$ is the observing wavelength.
(The qualitative explanation of Equation \ref{eq:ud} is that while
for unresolved sources the visibility is constant with increasing
uv radius, for progressively larger sources the visibility decreases 
faster with increasing uv radius.)
By comparing $V^2_{\rm calc}$ to the measured $V^2$ for a calibrator,
we derive the ``system visibility'', which represents the
point source response of the interferometer:
\begin{equation}
V^2_{\rm sys} = \frac{V^2_{\rm meas,calibrator}}{V^2_{\rm calc}}.
\end{equation}
This system visibility, in turn, is used to calibrate
the squared visibilities for the target source: 
\begin{equation}
V^2_{\rm target} = \frac{V^2_{\rm meas,target}}{V^2_{\rm sys}}.
\end{equation}
Specifically, we determine $V_{\rm sys}^2$ at the time of each 
target scan, using an average of $V_{\rm meas, calibrator}^2$ weighted by the
proximity of the target and calibrator in both time and angle.
For further discussion of the 
calibration procedure, see Boden et al. (1998).  

Calibrators must be close to the target sources (on the sky an in time)
so that the atmospheric effects will be the same for both.
They should also be of small
angular size, $\theta$, 
so that $V^2 \rightarrow 1$ and $dV^2_{\rm sys} / d\theta
\rightarrow 0$, and the calibration is thus
less sensitive to uncertainties in the assumed calibrator diameter.
The angular size of a calibrator can be estimated from the
published stellar luminosity and distance, from a blackbody fit to published
photometric data with the temperature constrained to
that expected for the published spectral type, or from 
an unconstrained blackbody fit to the photometric data.
We adopt the average of these three size estimates in our analysis,
and the uncertainty is given by the spread of these values.
Relevant properties of the calibrators used in these observations are
given in Table \ref{tab:cals}.

\section{Results \label{sec:res}}
We measured calibrated squared visibilities for AB Aur,
VV Ser, V1685 Cyg, AS 442,
and MWC 1080 (Table \ref{tab:obs}).  
All five sources are resolved by PTI (i.e., $V^2$ is significantly
different from unity), implying angular sizes
$\ga 1$ mas.
The data are consistent with disk-like morphologies for all sources, and
we can place good constraints on disk inclinations for most sources.
MWC 1080, V1685 Cyg, and VV Ser 
show evidence for significantly non-zero inclinations,
while a circularly symmetric distribution appears appropriate for AB Aur.
The AS 442 data are insufficient to constrain the inclination.

Interferometric observations of AB Aur and 
MWC 1080 at 2.2 $\mu$m have also been obtained with the
21-m and 38-m baselines of the IOTA interferometer \citep{M-G+99,M-G+01}.
When combined with our longer baseline PTI data (85-m and 110-m), 
these help fill in the $u-v$ plane, and enable 
us to improve constraints on source models  (see below; Figure 
\ref{fig:v1685cyg-model}). Based on discussion with R. Millan-Gabet,  
we assign an uncertainty to each IOTA visibility given by the standard
deviation of all data obtained for a given source with a given baseline.
We verify the registration of the IOTA and PTI data
using calibrators observed by both
interferometers.  Since the IOTA data for AB Aur was calibrated using
HD 32406, which is unresolved by both PTI and IOTA, we can be confident
of the registration.
We measured the diameter of HD 220074, the calibrator for MWC 1080,
to be $\theta_{\rm UD} = 1.98 \pm 0.06$, while MST
assumed a size of $2.10 \pm 0.22$.  This difference
in angular size translates into only a 0.7\% effect,
which is within the measurement errors (the effect is so small
because the calibrator is essentially unresolved by IOTA).

\subsection{Visibility Corrections \label{sec:viscorr}}
Nearby companions that lie outside the interferometric field of view, $\sim 50$
mas, but within the field of view of the detector, $\sim 1''$, will
contribute incoherent light to the visibilities. 
For MWC 1080, which has a known nearby companion \citep{CORPORON98},
we use the correction factor
\begin{equation}
\frac{V^2_{\rm true}}{V^2_{\rm meas}} = \left(\frac{1}
{1 + 10^{\Delta K / 2.5}}\right)^2,
\label{eq:companion}
\end{equation}
where $\Delta K$ is the difference in K-band magnitudes between
the two stars.  For MWC 1080, we measured $\Delta K = 2.70$ 
(angular separation $=0\rlap{.}''78$) using 
the Palomar Adaptive Optics system on the 200-inch telescope
on November 18, 2002.
V1685 Cyg is also known
to have a faint companion \citep[$\Delta K = 5.50$;][]{CORPORON98},
but the effect of this companion on the visibilities is negligible.
Adaptive optics images of the other sources in our sample show that 
none of these has any bright companions ($\Delta K < 5$) at distances between
$\sim 50$ mas and $1''$.  

Our measured visibilities contain information about
emission from both the circumstellar material and the star itself.
We can remove the effect of the central star
on the visibilities by including it in the models:
\begin{equation}
V^2_{\rm tot} = \left(\frac{F_{\ast} V_{\ast} + F_{\rm x} V_{\rm x}}{F_{\ast} +
F_{\rm x}} \right)^2 \approx 
\left(\frac{F_{\ast} + F_{\rm x} V_{\rm x}}{F_{\ast} +
F_{\rm x}} \right)^2,
\label{eq:vtot}
\end{equation}
where $F_{\ast}$ is the stellar flux, $F_{\rm x}$ is 
the excess flux (both
measured at 2.2 $\mu$m), $V_{\ast} \approx 1$ is the
visibility of the (unresolved) central star,
and $V_{\rm x}$ is the visibility due to the
circumstellar component.  It is reasonable to assume that $V_{\ast} \approx 1$,
since for typical stellar radii ($\sim 5$ R$_{\odot}$) and distances
($\sim 500$ pc), the angular diameters of the central stars will
be $\sim 0.5$ mas.
In the case of the binary model described below
(\S \ref{sec:binary}), we do not perform any such correction, since the
basic model already includes the stellar component.

Equation \ref{eq:vtot} assumes that the central star is a point source,
and thus contributes coherently to the visibilities. It is also possible that 
the starlight is actually observable only as scattered light emission, 
and that it will have some incoherent contribution to the visibility
($V_{\ast} \ne 1$).  
For example, coronographic
imaging with the Hubble Space Telescope has revealed scattered light
 on angular scales from $\sim 0\rlap{.}''1$--$9''$ around 
AB Aur \citep{GRADY+99}.  
A proper treatment of the effects of this scattered light 
on the visibilities is beyond the scope of this work, but we mention it
as a possible source of uncertainty.  Since the near-IR excess from
HAEBE sources typically
dominates over the near-IR stellar emission
(\S \ref{sec:phot}; Table \ref{tab:sources}),
the effect should be insignificant.

\subsection{Photometry \label{sec:phot}}
$F_{\ast}$ and $F_{\rm x}$ affect the visibilities (Equation
\ref{eq:vtot}), and thus it is important to determine these
quantities accurately.  Since HAEBE objects are often highly
variable at near-IR wavelengths
\citep[e.g.,][]{SKRUTSKIE+96}, we obtained photometric
K-band measurements of the sources in our sample that are
nearly contemporaneous with our PTI observations,
using the Palomar 200-inch telescope
between November 14 and 18, 2002.  Calibration relied on
observations of 2MASS sources close in angle to the target sources,
and we estimate uncertainties of $\sim 0.1$ magnitudes.
Our photometry is consistent with published measurements to within
$\sim 0.3$ magnitudes for all objects \citep{HILLENBRAND+92,EIROA+02}.

Following HSVK and MST, we calculate $F_{\ast}$ and $F_{\rm x}$
using our 
K-band photometry (Table \ref{tab:sources}) combined with BVRI photometry,
visual extinctions, and stellar effective temperatures from the literature
\citep{HILLENBRAND+92, OUDMAIJER+01, EIROA+02, BG70}.
De-reddening uses the extinction
law of Steenman \& Th\'{e} (1991).
Assuming that all of the short-wavelength
flux is due to the central star, we fit a blackbody at 
the assumed effective temperature to the de-reddened
BVRI data.  The K-band stellar flux
is derived from the value of this blackbody curve at 
2.2 $\mu$m.  The excess flux is then given by the difference between the
de-reddened observed flux and the stellar flux. The derived fluxes
are given in Table \ref{tab:sources}.
We note that VV Ser and AS 442 are optically variable by $\Delta V
\ga 1$ magnitudes on timescales of days to months \citep[while the
other sources in our sample show little or no optical variability;][]{HS99},
and thus $F_{\ast}$ is somewhat uncertain.  However,
since $F_{\rm x} / F_{\ast} \gg 1$ for these objects, this uncertainty
is negligible when modeling the visibilities.  

\subsection{Models \label{sec:mod}}
For each source, we compare the observed visibilities to those derived from
a uniform disk model, a Gaussian model, 
a ring model, and an accretion disk model with an inner disk hole
(all models are 2-D).  
If we assume that the inclination 
of the circumstellar material is zero, then the one remaining
free parameter in the
models is the angular size scale, $\theta$.  When we include inclination
effects, we fit for three parameters: size ($\theta$), 
inclination angle ($\phi$), and position angle ($\psi$).  
Inclination is defined such that a face-on disk has $\phi=0$,
and $\psi$ is measured east of north.  Following MST,
we include $\phi$ and $\psi$ in our models of the brightness distribution
via a simple coordinate transformation:
\begin{equation}
x' = x\sin\psi + y\cos\psi;\: y' = \frac{y\cos\psi - x\sin\psi}{\cos\phi}.
\end{equation}
Here, $(x,y)$ are the coordinates on the sky, and $(x',y')$ are the
transformed coordinates.  The effect of this coordinate 
transformation on the visibilities will be to transform $(u,v)$ to 
$(u',v')$:
\begin{equation}
u' = u \sin \psi + v \cos \psi;\:
v' = \cos \phi (v \sin \psi - u \cos \psi).
\end{equation}
Substitution of $(x',y')$ for $(x,y)$, and
$(u',v')$ for $(u,v)$ in the expressions below yield
models with inclination effects included. 

In addition to these four models, we also examine whether the
data are consistent with a wide binary model, which we approximate
with two stationary point sources.  For this model, the free parameters are
the angular separation ($\theta$), the position angle ($\psi$), and
the brightness ratio of the two components ($R$).

\subsubsection{Gaussian Model \label{sec:gauss}}
The brightness distribution for a normalized Gaussian model is given by
\begin{equation}
I_{\rm gauss}(x,y) = \exp \left(-\frac{4 \ln 2 \:\:
(x^2 + y^2)}{\theta^2}\right),
\end{equation}
and the (normalized) visibilities expected for this observed brightness 
distribution are calculated via a Fourier transform to be,
\begin{equation}
V_{\rm gauss}(r_{\rm uv}) = \exp \left(-\frac{\pi^2 \theta^2 r_{\rm uv}^2}
{4 \ln 2}\right).
\label{eq:gauss}
\end{equation}
Here, $(x,y)$ are the angular offsets from the central star, $\theta$ is
the angular FWHM of the brightness distribution,
and $r_{\rm uv} = (u^2+v^2)^{1/2}$ is the ``uv radius'' 
(Equation \ref{eq:ruv}).  The model for the observed
squared visibilities is obtained by using Equation \ref{eq:vtot} with
$V_{\rm x} = V_{\rm gauss}$.

\subsubsection{Uniform Disk Model \label{sec:uniform}}
The brightness distribution for a uniform disk is simply given by 
a 2-D top-hat function.  Thus, the normalized visibilities are given by
\begin{equation}
V_{\rm uniform}(r_{\rm uv}) = 2 \frac{J_1(\pi \theta r_{\rm uv})}{\pi \theta 
r_{\rm uv}}, 
\label{eq:uniform}
\end{equation}
where
$\theta$ is the angular diameter of the uniform disk brightness distribution,
and $r_{\rm uv} = (u^2+v^2)^{1/2}$ is the ``uv radius''
(Equation \ref{eq:ruv}).
The model for the observed squared 
visibilities is obtained by using Equation \ref{eq:vtot} with
$V_{\rm x} = V_{\rm uniform}$.

\subsubsection{Accretion Disk Model \label{sec:acc}}
We derive the brightness distribution and predicted visibilities
for a geometrically thin irradiated
accretion disk following the analysis of HSVK and
MST.  Assuming that the disk is heated by stellar 
radiation and accretion \citep{LBP74}, the temperature profile 
(in the regime where $R_{\ast}/R \ll 1$) is,
\begin{equation}
T_{\rm disk} = T_{\rm 1 AU} \left(\frac{R}{\rm AU}\right)^{-3/4},
\label{eq:acdisk1}
\end{equation}
where $T_{\rm 1 AU}$ is defined as the temperature at 1 AU, given by
\begin{equation}
T_{\rm 1 AU} = \left[ 2.52 \times 10^{-8} \left(\frac{R_{\ast}}
{R_{\odot}}\right)^3 T_{\ast}^4 + 
5.27 \times 10^{10} \left(\frac{M_{\ast}}{M_{\odot}}\right)
\left(\frac{\dot{M}}{10^{-5} \: M_{\odot} \: 
{\rm yr^{-1}}}\right)\right]^{1/4}.
\label{eq:t1au}
\end{equation}
We assume that the disk is truncated at an inner radius
$R_{\rm in}$, and an outer radius, $R_{\rm out}$.
Guided by Figure 14 of HSVK, we choose $R_{\rm in}$ to be the radius
where the temperature, $T_{\rm in}$, is 2000 K.
Thus, 
\begin{equation}
T_{\rm 1 AU} = 2000 \left(\frac{R_{\rm in}}{\rm AU}\right)^{3/4} .
\label{eq:trin}
\end{equation}
2000 K is a likely (upper limit) sublimation temperature
for the dust grains that make up circumstellar disks, and thus it is
reasonable that there would be little or no dust emission interior to
$R_{\rm in}$ (although the model does not exclude the possibility of optically
thin gas interior to $R_{\rm in}$).  
We choose $R_{\rm out}$ to be the lesser of 
1000 AU or the radius at which $T = 3$ K ($R_{\rm out}$ is not 
crucial in this analysis, since most of the near-IR flux comes
from the hotter inner regions of the disk).

The brightness distribution and visibilities for this disk are calculated
by determining the contributions from
a series of annuli from $R_{\rm in}$ to $R_{\rm out}$.
The flux in an annulus specified by inner boundary $R_{\rm i}$ and outer
boundary $R_{\rm f}$ is given by 
\begin{equation}
F_{\rm annulus} = \frac{\pi}{2 d^2} [B_{\nu}(T_{\rm i}) + B_{\nu}(T_{\rm f})]
(R_{\rm f}^2 - R_{\rm i}^2),
\end{equation}
and the visibilities for this annulus are
\citep[following][]{M-G+01},
\begin{equation}
V_{\rm annulus} = \frac{ \pi}{d^2} [B_{\nu}(T_{\rm i}) + B_{\nu}(T_{\rm f})]
\left[R_{\rm f}^2 
\frac{J_1(\pi \theta_{\rm f} r_{\rm uv})}{\pi \theta_{\rm f} r_{\rm uv}} - 
R_{\rm i}^2 
\frac{J_1(\pi \theta_{\rm i} r_{\rm uv})}
{\pi \theta_{\rm i} r_{\rm uv}}\right].
\end{equation}
Here, $d$ is the distance to the source, $\nu$ is the observed frequency,
$B_{\nu}$ is the Planck function,
$T$ is the temperature, $R$ is the physical radius,
$\theta$ is the 
angular size, $r_{\rm uv}$ is the ``uv radius'' (Equation \ref{eq:ruv}), 
and $i,f$ indicate the inner
and outer boundaries of the annulus.
To obtain the visibilities for the entire disk, we
sum the visibilities for each annulus, and normalize by the total flux:
\begin{equation}
V_{\rm disk} = \frac{\sum_{R_{in}}^{R_{out}} V_{\rm annulus}}{
\sum_{R_{in}}^{R_{out}}{F_{\rm annulus}}}.
\end{equation}
The resultant model visibilities are obtained by plugging this expression
into Equation \ref{eq:vtot}.  We note that although we do not use
the observed excess K-band flux to constrain the disk model, 
we do verify that the total flux in the model is consistent
(to within a factor of 2) with the observations.

\subsubsection{Ring Model \label{sec:ring}}
The brightness distribution for a uniform ring model is given by
\begin{equation}
I_{\rm ring}(x,y) = \cases{\rm constant & if $\frac{\theta_{\rm in}}{2} < 
\sqrt{x^2+y^2} < \frac{\theta_{\rm out}}{2}$ \cr 0
& otherwise}.
\end{equation}
Here, $(x,y)$ are the angular offsets from the central star.
We define the width of the ring via the relation $f = W/R$, where $R$ is the
radius of the inner edge of the ring, and $W$ is the width of the ring.
Using this relation, we write the inner and outer angular radii of the
ring as $\theta_{\rm in}$ and $\theta_{\rm out} = (1+f)\theta_{\rm in}$.
The normalized visibility of the ring is given by
\begin{equation}
V_{\rm ring} = \frac{2}{\pi \theta_1 (2f + f^2)} \left[
(1+f) J_1([1+f] \pi \theta_{\rm in} r_{\rm uv}) - 
J_1(\pi \theta_{\rm in} r_{\rm uv}) \right],
\label{eq:ring}
\end{equation}
where $r_{\rm uv} = (u^2 + v^2)^{1/2}$ is the ``uv radius''
(Equation \ref{eq:ruv}).
The model for the observed
visibilities is obtained by using Equation \ref{eq:vtot} with
$V_{\rm x} = V_{\rm ring}$.

In order to facilitate comparison of our data to puffed up inner disk
models from the literature, we will use ring widths derived from
radiative transfer modeling by DDN.  Specifically,
for stars earlier than spectral type B6, we assume $f = 0.27$, 
and for stars later than B6, we assume $f=0.18$ \citep[Table 1 from][]{DDN01}.

\subsubsection{Two-Component Model \label{sec:binary}}
This model simulates a wide binary, where 
visibilities are effectively due to two stationary point sources, 
with some flux
ratio and angular separation vector.
We explore flux ratios from 0.2 to 1, and angular separations from 1 to
100 mas.  For flux ratios $<0.2$, or angular separations $<1$ mas,
the effects of the companions on the visibilities will be negligible,
and we can rule out angular separations $\ga 100$ mas from adaptive optics
imaging (\S \ref{sec:viscorr}).  
The squared visibility for the binary model is,
\begin{equation}
V^2_{\rm binary} = \frac{1 + R^2 + 2R \cos\left(\frac{2\pi}{\lambda}
\vec{B} \cdot \vec{s} \right)}{(1+R)^2},
\label{eq:binary}
\end{equation}
where $(\vec{B} \cdot \vec{s})/\lambda = \theta [u \sin(\psi) + v \cos(\psi)]$,
$\theta$ is the angular separation of the binary, $\psi$ is the
position angle, $R$ is the ratio of the fluxes of the two components,
and $\lambda$ is the observed wavelength.

\subsection{Modeling of Individual Sources}
For each source, we fit the PTI and IOTA visibility data with the models
described in \S \ref{sec:mod} using grids of parameter values.
The grid for face-on disk models was generated
by varying $\theta$ from 0.01 to 10 mas
in increments of 0.01 mas.  For inclined disk models, 
in addition to varying $\theta$, we varied
$\phi$ from 0$^{\circ}$ to $90^{\circ}$ and
$\psi$ from 0$^{\circ}$ to $180^{\circ}$, both in increments of $1^{\circ}$.
As mentioned above, 
$\phi=0$ corresponds to face-on, and $\psi$ is measured east of north.
Since inclined disk models are symmetric under
reflections through the origin, we do not explore
position angles between $180^{\circ}$ and $360^{\circ}$.
For the binary model, we varied $\theta$ from 1 to 100 mas
in increments of 0.01 mas, $\psi$ from 0$^{\circ}$ to $180^{\circ}$ 
in increments of $1^{\circ}$, and $R=F_2/F_1$ from 0.2 to 1 in increments
of 0.001.

For each point in the parameter grid, we generated
a model for the observed $u-v$ coverage, and calculated the
reduced chi squared ($\chi_r^2$) to determine the
``best-fit'' model.  1-$\sigma$ confidence limits were determined
by finding the grid points where $\chi_r^2$ equals the minimum value 
plus one. For inclined disk or binary models,
the confidence limits on each parameter were determined by projecting
the 3-D $\chi_r^2$ = min$+1$ surface onto the 1-D parameter spaces.

\begin{figure}
\plotone{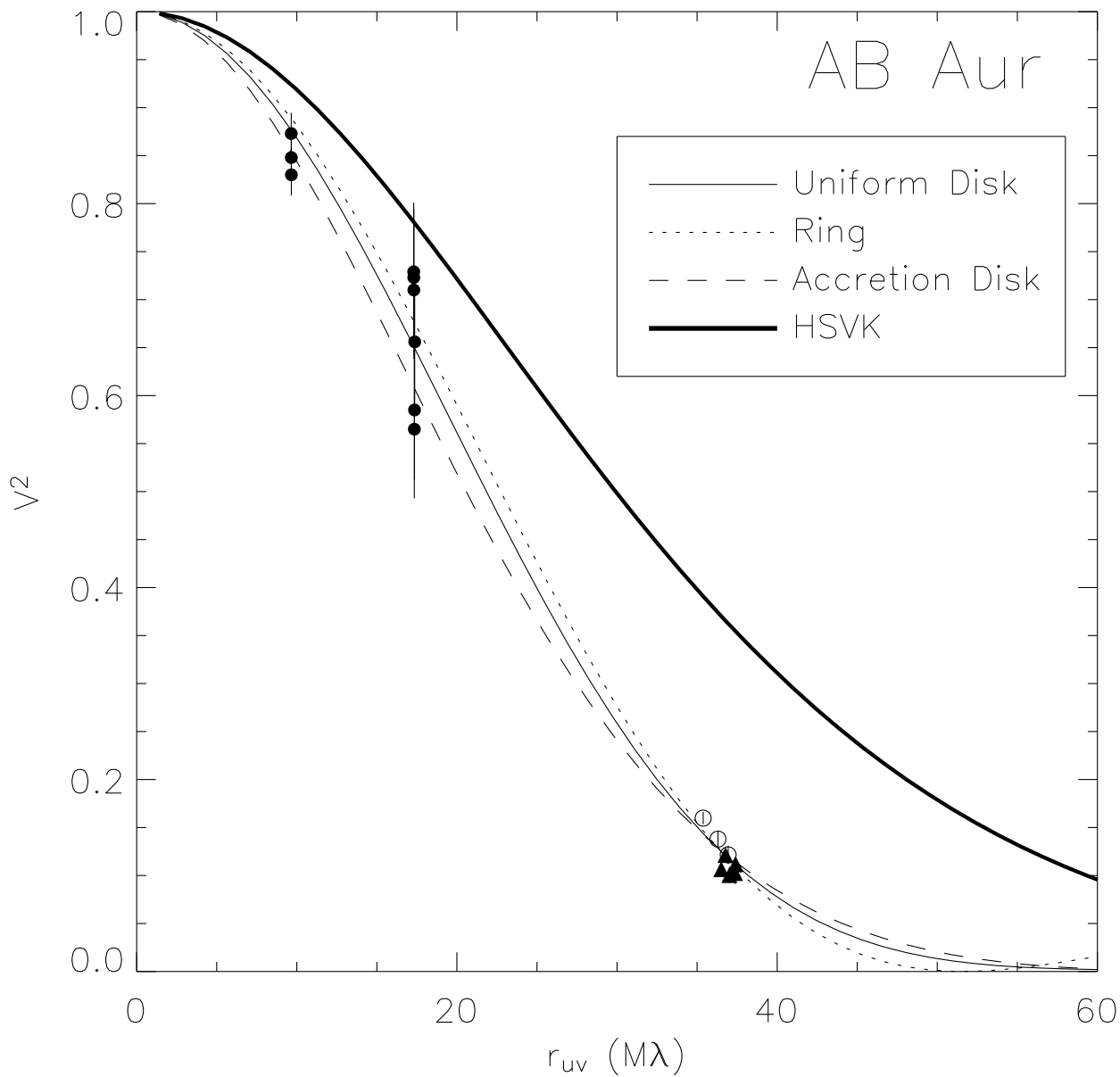}
\caption{$V^2$ data from PTI  (symbols) and IOTA (filled dots; MST)
for AB Aur, as a function of $r_{\rm uv} = (u^2 + v^2)^{1/2}$.  
PTI data for individual nights are represented by different symbols.
Face-on
uniform disk (solid line), ring (dotted line), and accretion disk
(dashed line) models are over-plotted.  
We also plot the visibilities calculated for an
accretion disk model with $R_{\rm in} = 0.09$
AU and $T_{\rm in} = 2360$ K (HSVK; thick solid line).  
\label{fig:abaur-uv}}
\end{figure}

\begin{figure}
\plotone{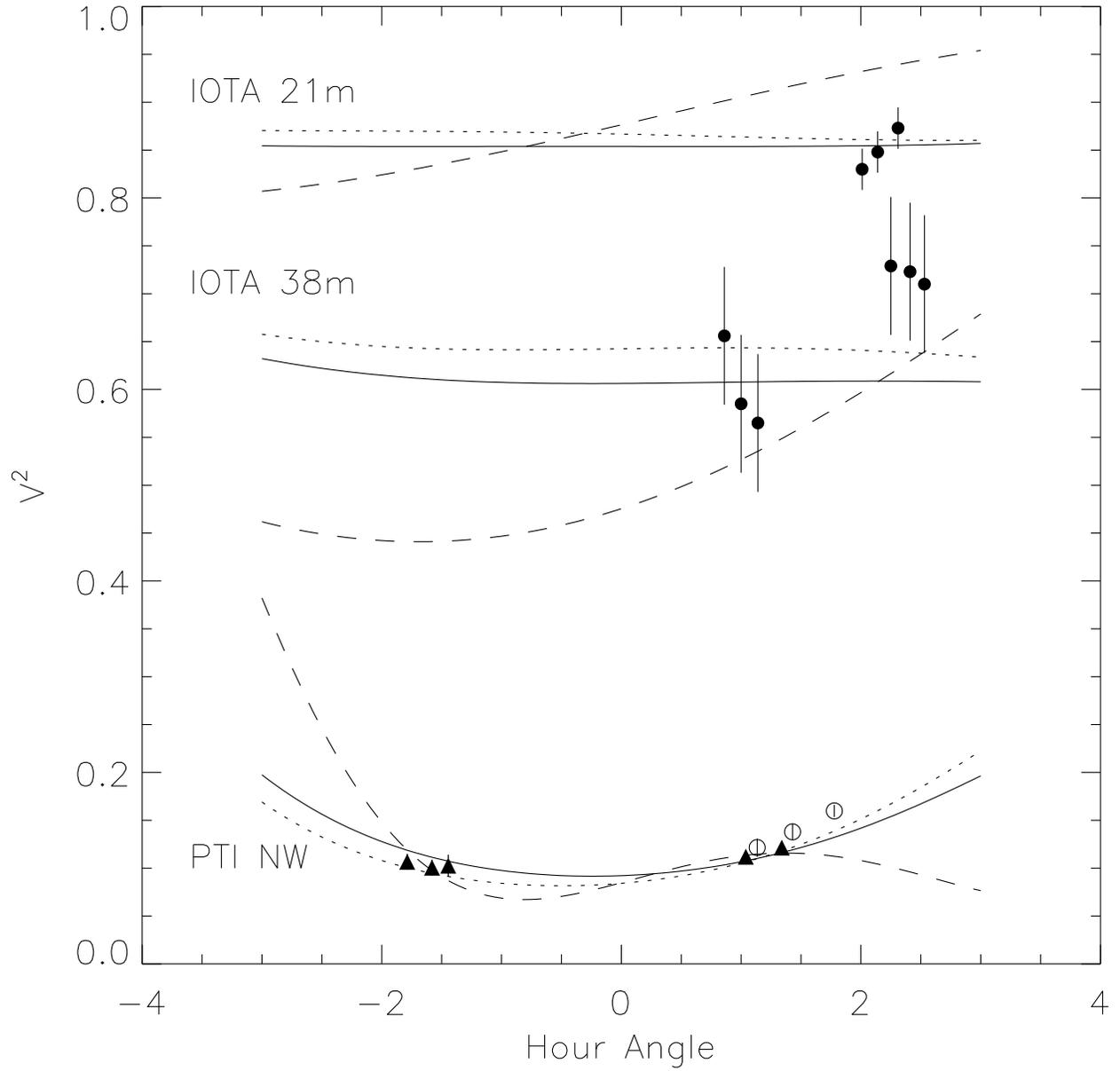}
\caption{PTI and IOTA $V^2$ data 
for AB Aur (represented as in Figure \ref{fig:abaur-uv}), 
as a function of hour angle. 
Over-plotted are face-on and inclined accretion disk models
(solid and dotted lines, respectively),
as well as the best-fit binary model (dashed line).
\label{fig:abaur-ha}}
\end{figure}

\begin{figure}
\plotone{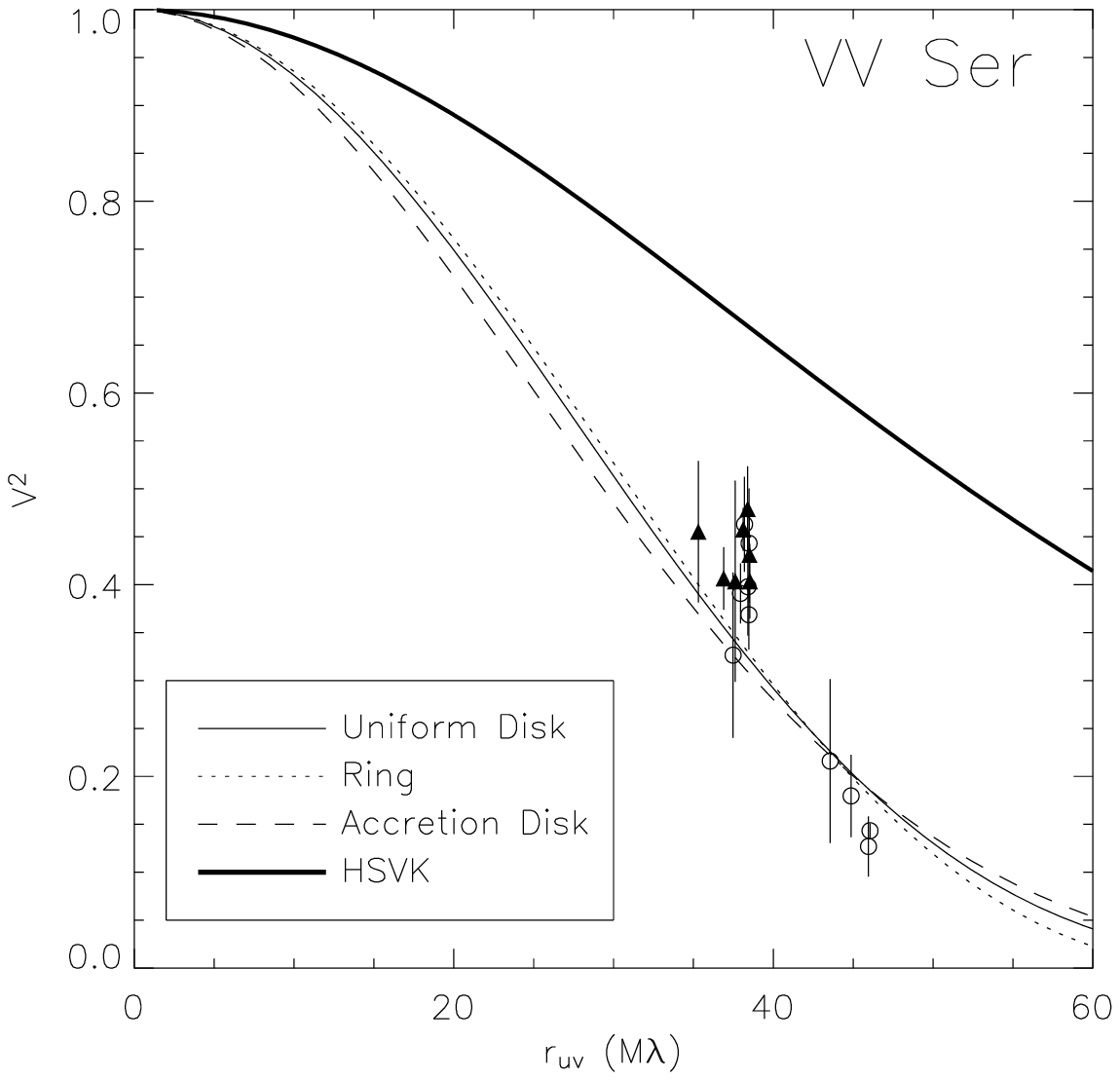}
\caption{PTI $V^2$ data for VV Ser, as a function of
$r_{\rm uv} = (u^2 + v^2)^{1/2}$. 
PTI data for individual nights are represented by different symbols.
Face-on
uniform disk (solid line), ring (dotted line), and accretion disk
(dashed line) models are over-plotted.  
We also plot the visibilities calculated for an
accretion disk model with $R_{\rm in} = 0.08$ AU and $T_{\rm in} = 2710$ K
(HSVK; thick solid line).  
\label{fig:vvser-uv}}
\end{figure}

Tables \ref{tab:abaur}-\ref{tab:mwc1080} list the best-fit angular size scales
($\theta$)
for face-on models, the sizes ($\theta$), position angles ($\psi$), 
and inclinations ($\phi$) for inclined disk models, and the angular 
separations ($\theta$), position angles ($\psi$), 
and brightness ratios ($R$) for binary models.  Values of $\chi_r^2$ are
also included in the Tables.
Figures \ref{fig:abaur-uv}, \ref{fig:vvser-uv}, \ref{fig:v1685cyg-uv},
\ref{fig:as442-uv}, and \ref{fig:mwc1080-uv} 
show plots of observed $V^2$ versus $r_{\rm uv}$
for each source along with the curves predicted by various face-on models.
Inclined models are not circularly symmetric, and 
the visibilities are a function of the observed
position angle in addition to the projected baseline 
(Figure \ref{fig:v1685cyg-model}).
We plot the observed and modeled $V^2$ for inclined models as
a function of hour angle in Figures \ref{fig:abaur-ha}, 
\ref{fig:vvser-ha}, \ref{fig:v1685cyg-ha},
\ref{fig:as442-ha}, and \ref{fig:mwc1080-ha}.

The best-fit binary separations for all sources in our sample are
$\ga 2.5$ mas.  For the distances and approximate masses of the sources in our 
sample, these separations correspond to orbital periods of many years.
Thus, our assumption that the two point sources in the binary model
are stationary is justified.

\subsubsection{AB Aur}
The visibilities for AB Aur are consistent with a disk-like circumstellar 
distribution that is inclined by $\la 30^{\circ}$
(Figures \ref{fig:abaur-uv}--\ref{fig:abaur-ha}).
From Table \ref{tab:abaur},  
the best-fit models indicate size 
scales\footnote{ As outlined in \S \ref{sec:gauss}--\ref{sec:ring},
characteristic size scales for different models measure
different parts of the brightness distributions: 
Gaussian models measure full widths at half
maxima, uniform disk models measure outer diameters, accretion disk models 
measure inner disk diameters, and ring models measure inner ring diameters.
The spread in quoted angular sizes for a source is 
mainly due to these differences.} between 2.2 and 5.8 
mas,
and an inclination angle between $27^{\circ}$ and 35$^{\circ}$.
The values of $\chi_r^2$ are significantly
lower for inclined models than for face-on models
($\chi_r^2 \sim 1$ and $2$, respectively; Table \ref{tab:abaur}), 
and the data cannot be fit well by a binary model ($\chi_r^2 \sim 8$).

\subsubsection{VV Ser}
\begin{figure}
\plotone{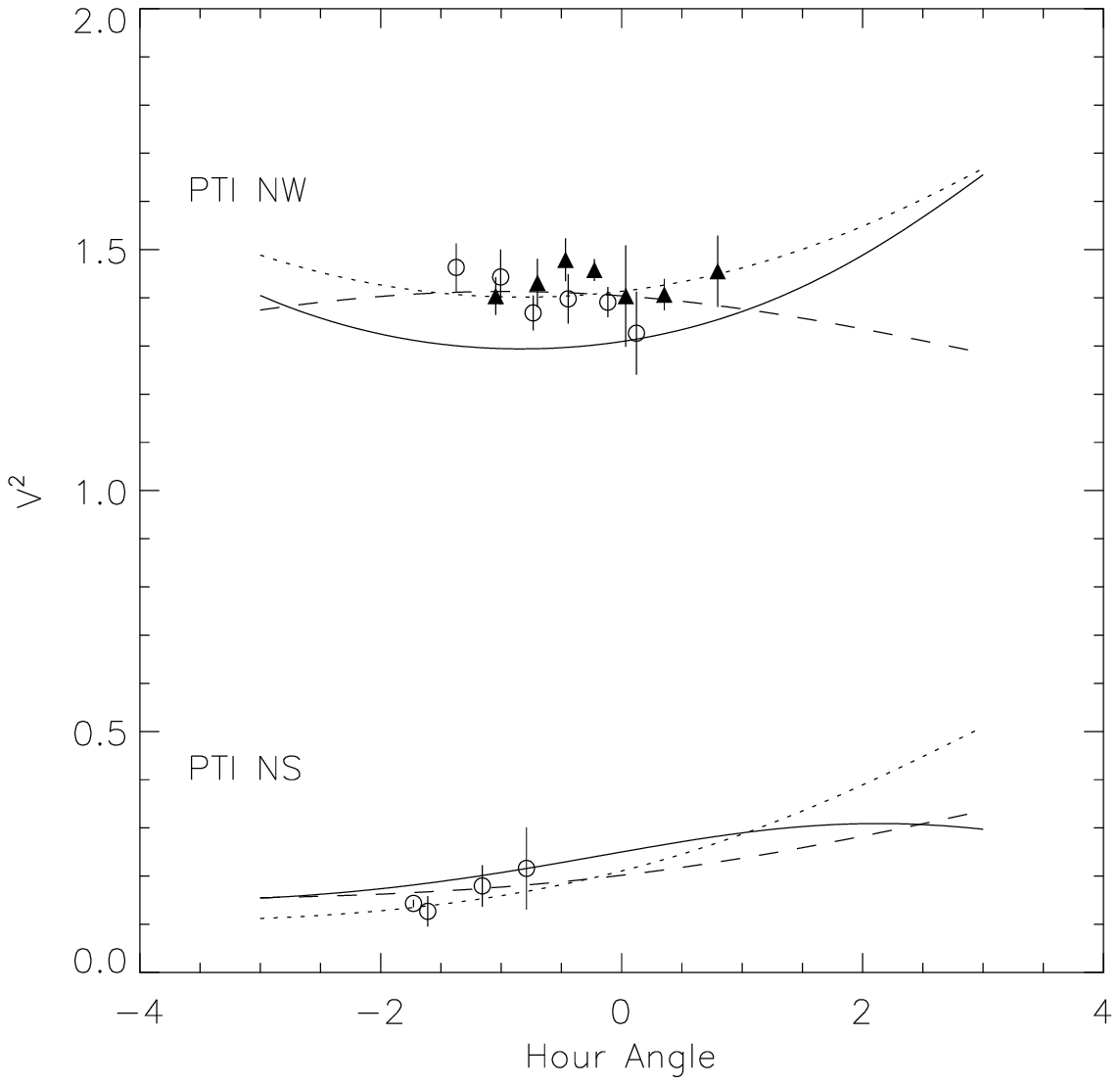}
\caption{PTI $V^2$ data for VV Ser (represented as in Figure 
\ref{fig:vvser-uv}), as a function of hour angle. 
For clarity, we have plotted $V^2 + 1$ for the data
taken with the NW baseline.
Over-plotted are face-on and inclined accretion disk models
(solid and dotted lines, respectively),
as well as the best-fit binary model (dashed line).
\label{fig:vvser-ha}}
\end{figure}

The angular size scales for best-fit disk
models range from 1.5 to 3.9, and the disk inclinations are between
$80^{\circ}$ and $90^{\circ}$ (Table \ref{tab:vvser}).
An inclined disk model clearly fits
the VV Ser data better than a face-on model (Figures \ref{fig:vvser-uv} and
\ref{fig:vvser-ha}). 
Inclined model fits give $\chi_r^2 < 1$, while face-on model fits
have $\chi_r^2 > 5$ (Table \ref{tab:vvser}). 
However, as indicated in Figure \ref{fig:v1685cyg-model},
the $u-v$ coverage for this object is rather sparse, and precludes
placing stringent constraints on the value of $\phi$.
Moreover, with such sparse $u-v$ coverage a binary model cannot be
ruled out (Figure \ref{fig:vvser-ha}).

\subsubsection{V1685 Cyg}
The size scales for V1685 Cyg under the assumptions of various disk models
range from 1.3 to 3.9 mas, and the inclinations are between
49$^{\circ}$ and 51$^{\circ}$ (Table \ref{tab:v1685cyg}).
The visibility data are not fit very well by any model,
although of those considered, inclined disks fit best
(Figures \ref{fig:v1685cyg-uv}--\ref{fig:v1685cyg-ha}).  
While we cannot rule out a binary model, we note that $\chi_r^2 \sim 3$ 
for the binary model, compared to $\chi_r^2 \sim 2$ for inclined disk models
(Table \ref{tab:v1685cyg}).
Better coverage of the  $u-v$ plane
should help to improve our understanding of this source (Figure 
\ref{fig:v1685cyg-model}).

\subsubsection{AS 442}
The PTI data for AS 442 generally have low signal-to-noise, and it is 
difficult to distinguish between different models.  Nevertheless,
we can make an approximate determination of the size scale, although we
cannot distinguish between inclined disk, face-on disk, or binary models
(Figures \ref{fig:as442-uv} and \ref{fig:as442-ha}). 
The size scales for various disk models range from
0.9 to 2.7 mas (Table \ref{tab:as442}).

\subsubsection{MWC 1080}
The PTI and IOTA observations for MWC 1080 are completely incompatible
with face-on models ($\chi_r^2 > 40$), 
and significantly non-zero inclinations are required to fit the
data well (Figures \ref{fig:mwc1080-uv} and \ref{fig:mwc1080-ha}).  
The best-fit inclination angles for various disk models
range from $51^{\circ}$ to $56^{\circ}$, and the angular size scales
are between 1.5 and 4.1 mas (Table \ref{tab:mwc1080}).
For this source, we can rule out a binary model with a relatively high degree 
of confidence: $\chi_r^2 \sim 10$ for the binary model, compared to
$\chi_r^2 \sim 2$ for inclined disk models.

\section{Discussion \label{sec:disc}}
As discussed in \S \ref{sec:intro}, there is currently a wide variety
of evidence that supports the existence of circumstellar disks around many
HAEBE stars.  Our new PTI results strengthen this contention.  
Resolved, small-scale ($\sim 1$ AU)
distributions of dust are found in all observed sources, and the 
non-symmetric intensity distributions of best-fit models for most objects
provide support for inclined disk geometries.


We suggest that the 
material around VV Ser, V1685 Cyg, and MWC 1080 is significantly inclined,
and we cannot rule out a high inclination angle for AS 442.
This hypothesis is compatible with observed optical variability in
VV Ser and AS 442 
\citep[$\Delta V_{\rm VV ser} \sim 2$, $\Delta V_{\rm AS 442} \sim 1$;][]
{HS99}, which has been attributed to variable obscuration from clumps of
dust orbiting in inclined circumstellar disks.
The AB Aur data, 
in contrast, are consistent with a circumstellar distribution
that is within $35^{\circ}$  of face-on.  This agrees well with
MST and is compatible with modeling of scattered 
light observed with the Hubble Space Telescope, which suggests an inclination
angle $\la 45^{\circ}$ \citep{GRADY+99}.  The small 
amplitude of variability in AB Aur \citep[$\Delta V \sim 0.25$;][]{HS99}
is also consistent with this low inclination angle (under the assumption
that variability is caused by time-dependent circumstellar obscuration).
The low inclination angle does {\it not}, however, agree with 
mm-wave imaging in the $^{13}$CO(1-0) line, which 
yields an estimated inclination of $76^{\circ}$ for the AB Aur disk 
\citep{MS97}.



The angular sizes determined from our observations
are generally in good agreement with
the non-inclined ($\phi=0$) 
flat accretion disk models of HSVK
for early-type Herbig Be stars, V1685 Cyg and MWC 1080, but not for
the later-type stars, AB Aur, VV Ser, and AS 442
(the spectral type for VV Ser, A0, is uncertain by $\pm 5$
spectral subclasses; Mora et al. 2001).
Angular sizes derived from the earlier IOTA
observations (MST) were often an order of magnitude larger than those 
predicted by the HSVK models, and on this basis MST ruled these models out. 

\begin{figure}
\plotone{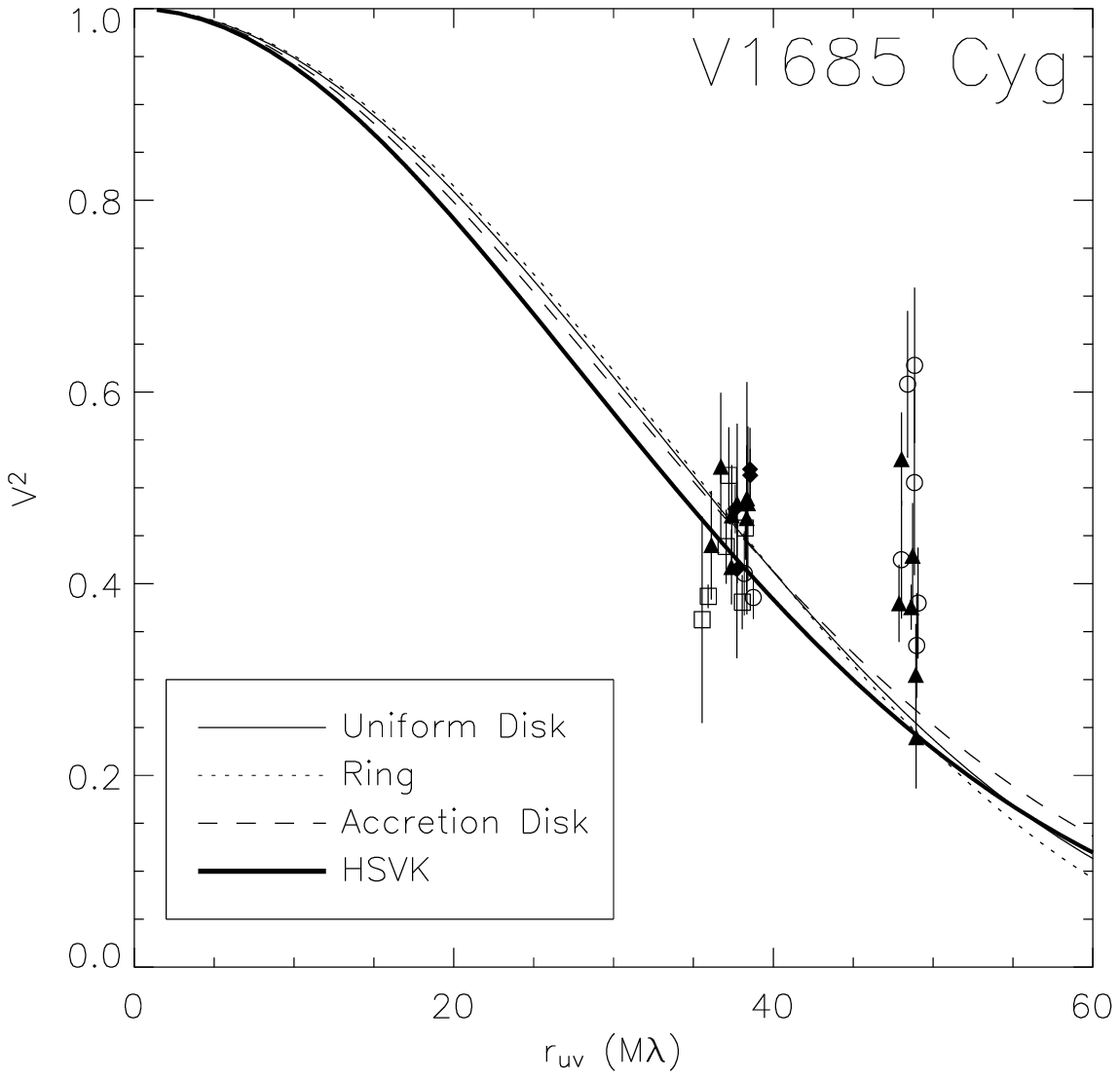}
\caption{PTI $V^2$ data for V1685 Cyg, as a function of
$r_{\rm uv} = (u^2 + v^2)^{1/2}$. 
PTI data for individual nights are represented by different symbols.
Face-on
uniform disk (solid line), ring (dotted line), and accretion disk
(dashed line) models are over-plotted.  
We also plot the visibilities calculated for an
accretion disk model with $R_{\rm in} = 0.63$ AU and $T_{\rm in} = 2060$ K
(HSVK; thick solid line).  
\label{fig:v1685cyg-uv}}
\end{figure}

\begin{figure}
\plotone{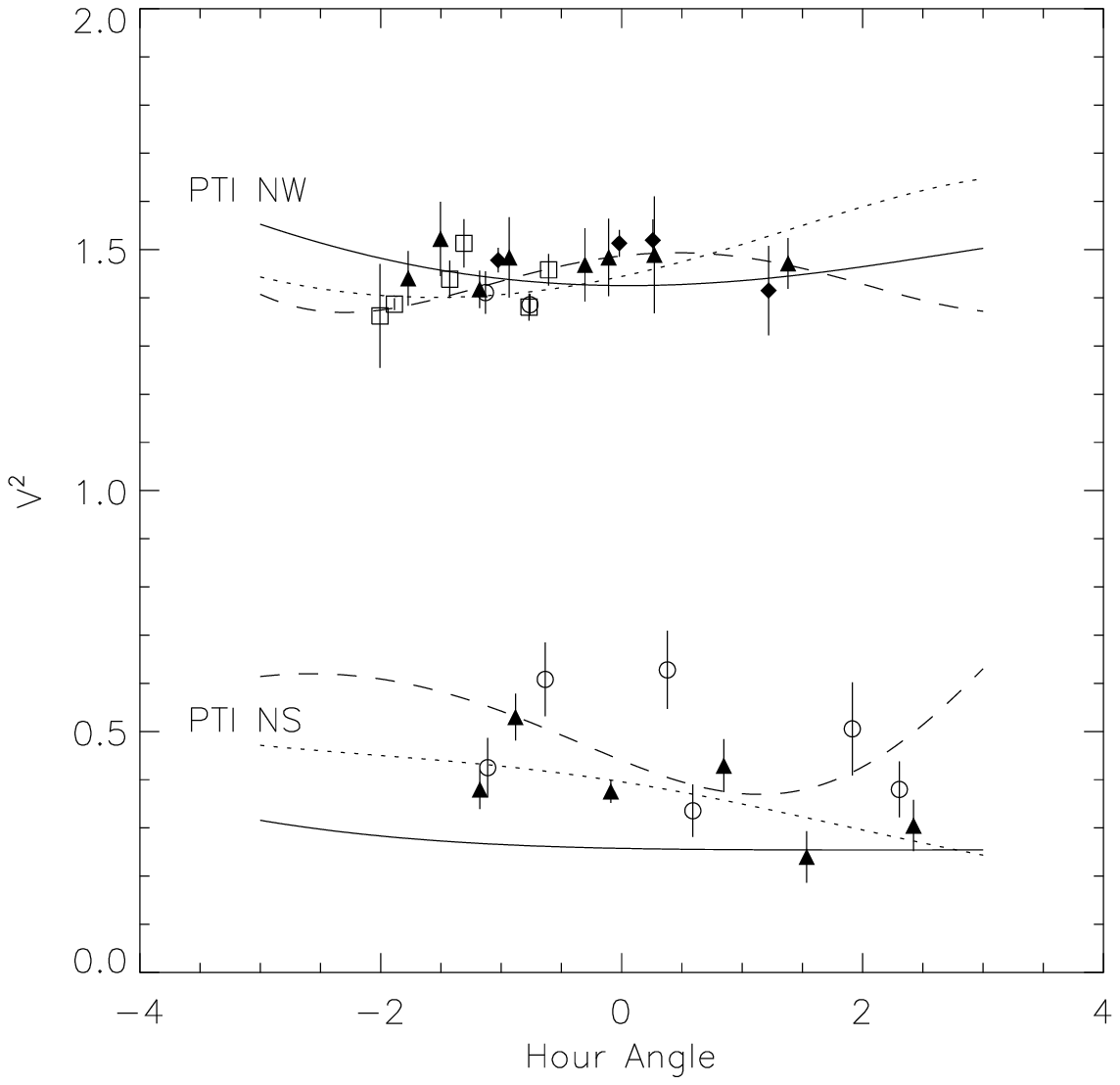}
\caption{PTI $V^2$ data for V1685 Cyg (represented as in Figure 
\ref{fig:v1685cyg-uv}), as a function of hour angle. 
For clarity, we have plotted $V^2 + 1$ for the data
taken with the NW baseline.
Over-plotted are face-on and inclined accretion disk models
(solid and dotted lines, respectively),
as well as the best-fit binary model (dashed line).
\label{fig:v1685cyg-ha}}
\end{figure}

\begin{figure}
\plotone{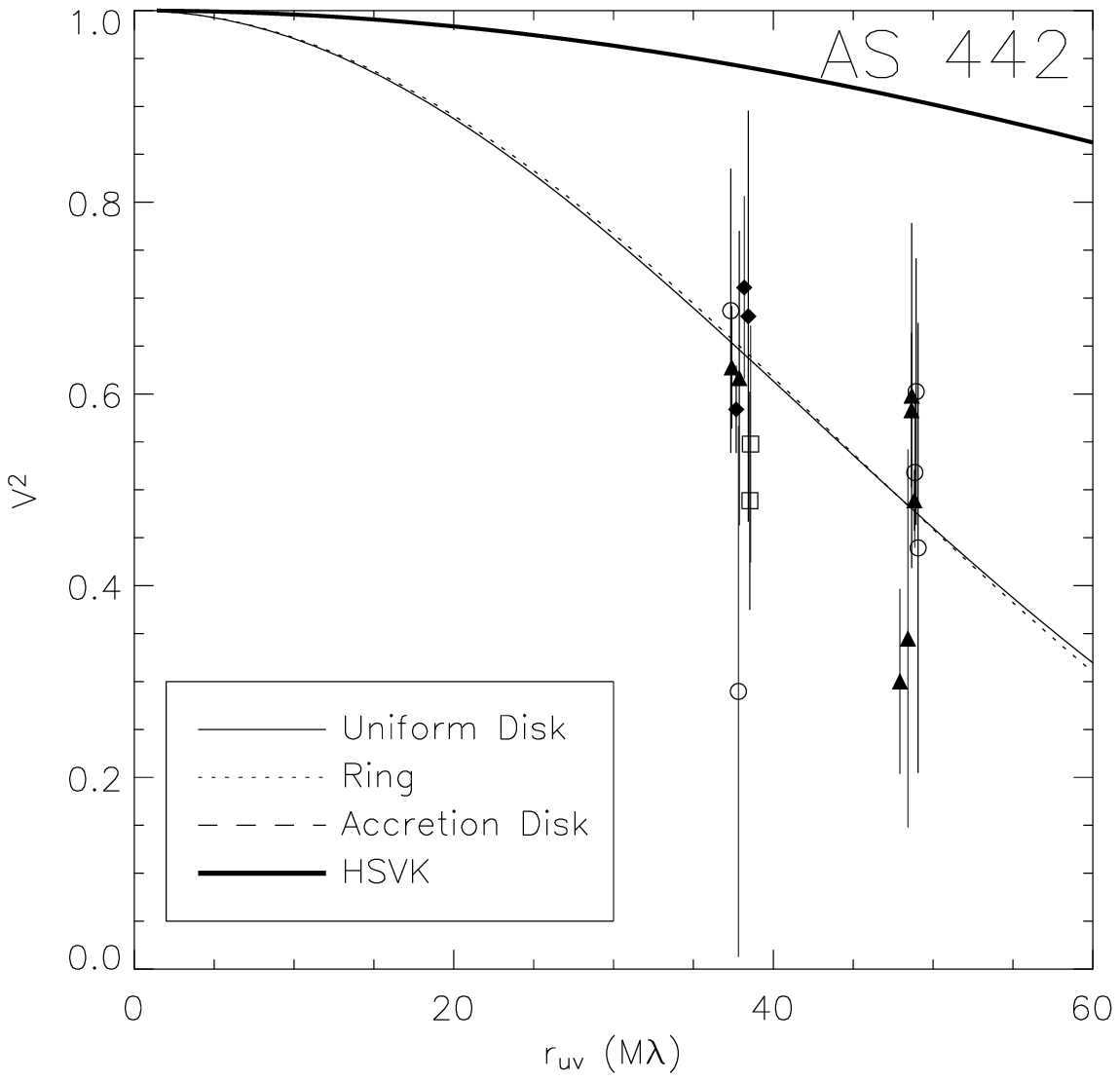}
\caption{PTI $V^2$ data for AS 442, as a function of
$r_{\rm uv} = (u^2 + v^2)^{1/2}$. 
PTI data for individual nights are represented by different symbols.
Face-on
uniform disk (solid line), ring (dotted line), and accretion disk
(dashed line) models are over-plotted.  
We also plot the visibilities calculated for an
accretion disk model with $R_{\rm in} = 0.10$ AU and $T_{\rm in} = 2000$ K
(HSVK; thick solid line).  
\label{fig:as442-uv}}
\end{figure}

\begin{figure}
\plotone{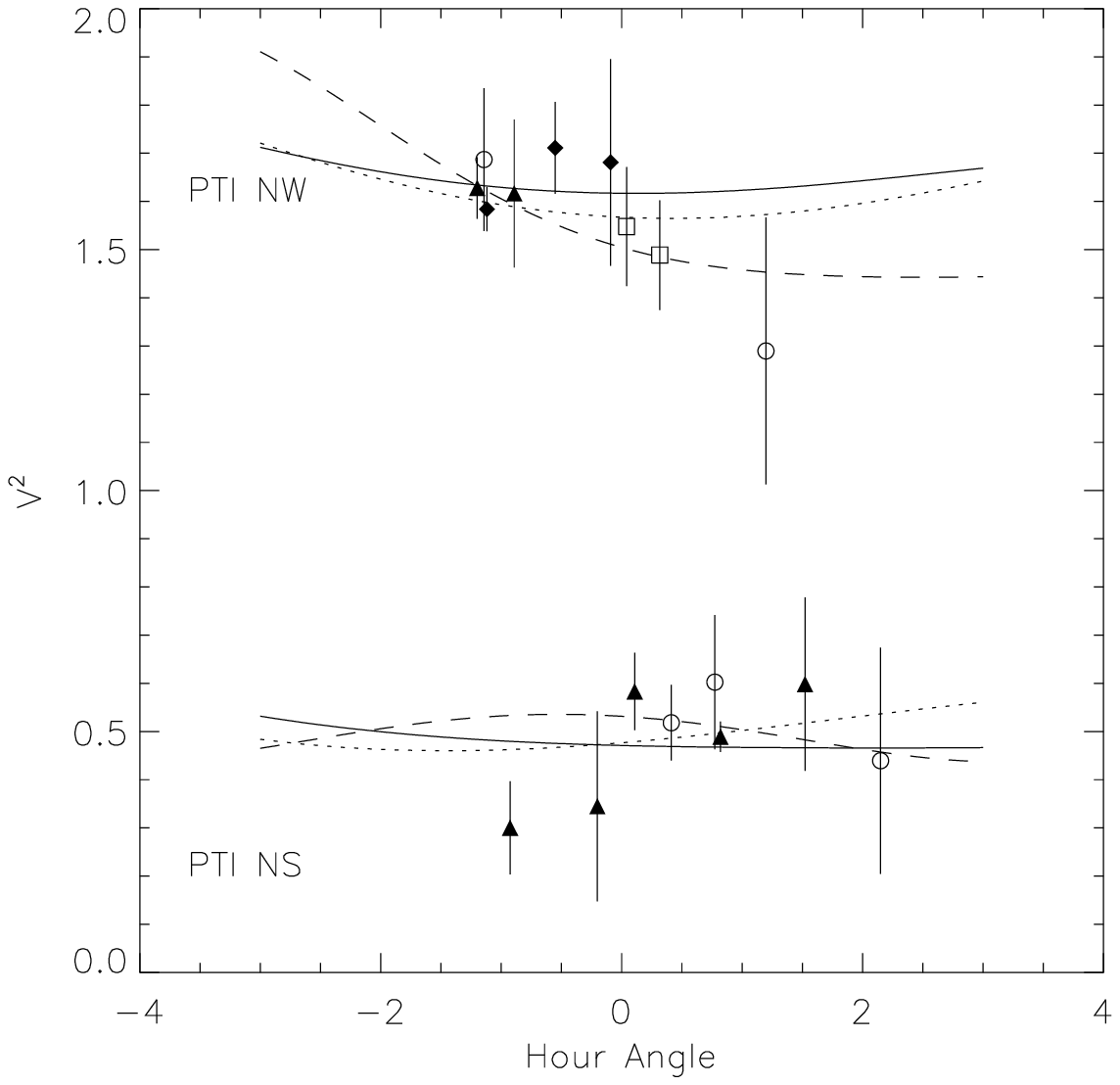}
\caption{PTI $V^2$ data for AS 442 (represented as in Figure 
\ref{fig:as442-uv}), as a function of hour angle. 
For clarity, we have plotted $V^2 + 1$ for the data
taken with the NW baseline.
Over-plotted are face-on and inclined accretion disk models
(solid and dotted lines, respectively),
as well as the best-fit binary model (dashed line).
\label{fig:as442-ha}}
\end{figure}

\begin{figure}
\plotone{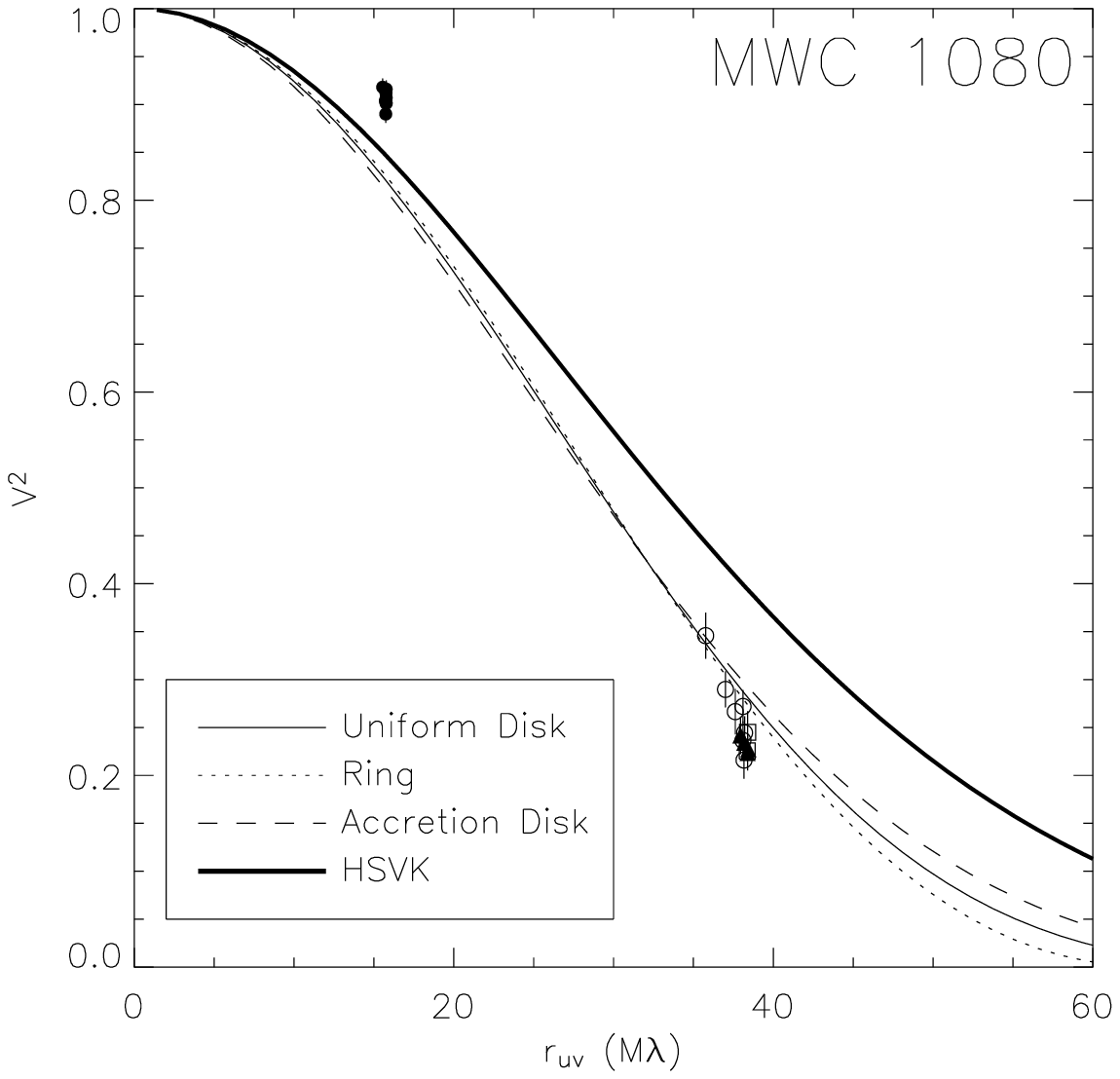}
\caption{$V^2$ data from PTI  (symbols) and IOTA (filled dots; MST)
for MWC 1080, as a function of $r_{\rm uv} = (u^2 + v^2)^{1/2}$.  
PTI data for individual nights are represented by different symbols.
Face-on
uniform disk (solid line), ring (dotted line), and accretion disk
(dashed line) models are over-plotted.  
We also plot the visibilities calculated for an
accretion disk model with $R_{\rm in} = 0.59$
AU and $T_{\rm in} = 2490$ K (HSVK; thick solid line).  
While none of these face-on models fit the data well, good fits
are obtained with inclined models (Figure \ref{fig:mwc1080-ha}).
\label{fig:mwc1080-uv}}
\end{figure}

\begin{figure}
\plotone{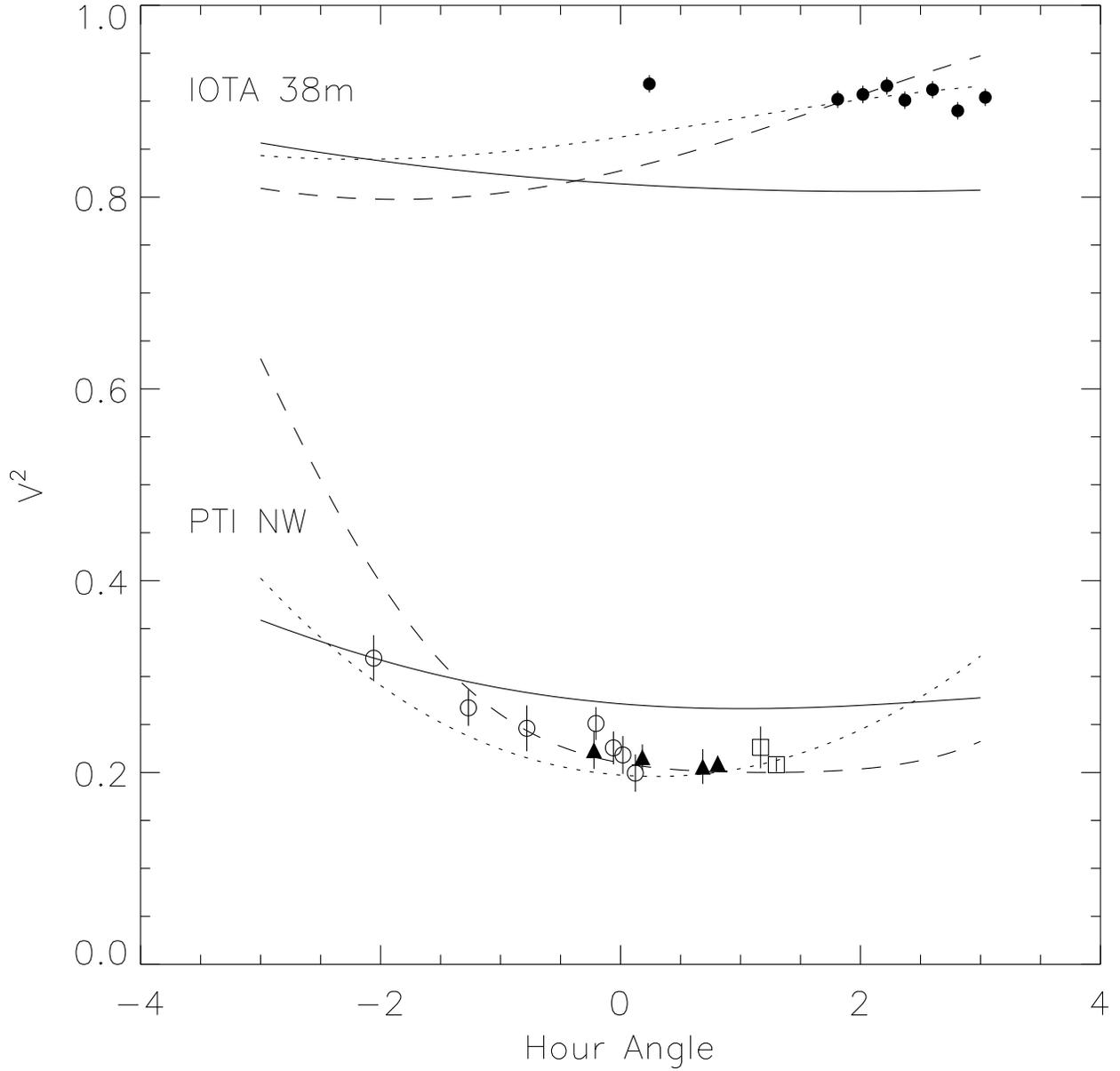}
\caption{PTI and IOTA $V^2$ data 
for MWC 1080 (represented as in Figure \ref{fig:mwc1080-uv}), 
as a function of hour angle. 
Over-plotted are face-on and inclined accretion disk models
(solid and dotted lines, respectively),
as well as the best-fit binary model (dashed line).
Note the significant improvement in the fit when inclination effects
are included in the model.
\label{fig:mwc1080-ha}}
\end{figure}

HSVK determined the best-fit models for the SEDs
of HAEBE sources by assuming a face-on disk geometry,
adjusting the accretion rate to match the mid-IR flux,
and then adjusting the size of the inner hole to match the near-IR flux.
We compare our results with theirs in a
qualitative way by
plotting the visibilities predicted by the HSVK
models along with the observed PTI and IOTA visibilities
in Figures \ref{fig:abaur-uv}, \ref{fig:vvser-uv}, \ref{fig:v1685cyg-uv},
\ref{fig:as442-uv}, and \ref{fig:mwc1080-uv}.
For a more quantitative comparison, we use published luminosities,
effective temperatures, and accretion rates \citep{HILLENBRAND+92} to calculate
the inner radii predicted by flat accretion disk
models with $T_{\rm in}=2000$ K (Equations
\ref{eq:t1au} and \ref{eq:trin}), 
and compare these estimates to our interferometric results
(which were also derived assuming $T_{\rm in}=2000$ K; \S \ref{sec:acc}).
In Table \ref{tab:acc}, $R_{\rm face-on}$ and 
$R_{\rm inclined}$
are the inner radii determined by fitting the interferometric data to
face-on and inclined
accretion disk models, respectively (\S \ref{sec:acc}), and $R_{\dot{M}=0}$,
$R_{\dot{M} \ne 0}$ are the radii calculated using the HSVK flat disk models 
without accretion, and with accretion effects included, respectively.
No estimate of $\dot{M}$ is available for AS 442.

Our data for the later-type stars AB Aur, VV Ser, and AS 442 are
fairly consistent with the puffed up inner disk models of DDN,
assuming inner disk temperatures $\ga 2000$ K.  In contrast,
puffed up inner disk models are completely incompatible with
the PTI results for the very early-type stars in our sample, 
V1685 Cyg and MWC 1080.  
The radius of the inner wall, $R_{\rm in}$, predicted
by DDN is,
\begin{equation}
R_{\rm in} = \sqrt{\frac{L_{\ast}}{4\pi T_{\rm in}^4 \sigma} (1+f)},
\end{equation}
where, $L_{\ast}$ is the (published) stellar luminosity,
$T_{\rm in}$ is the temperature of the inner wall,  
and $f$ is the ratio of the width of the inner wall to its radius.
Based on DDN, we assume $f=0.27$ for stars earlier
than spectral type B6, and $f=0.18$ for later-type stars.
We calculate $R_{\rm in}$ for $T_{\rm in} = 1500, 2000$ K
(likely sublimation temperatures for silicate and graphite dust
grains, respectively), and
compare these to the ring diameters derived from fitting to near-IR
interferometric visibilities. In  Table \ref{tab:ring},
$R_{\rm face-on}$ and $R_{\rm inclined}$ represent the inner radii 
determined for face-on and inclined ring models, respectively 
(\S \ref{sec:ring}), and $R_{2000}$, $R_{1500}$ are the inner radii 
predicted by the DDN puffed-up inner disk models,
assuming sublimation temperatures of 2000 and 1500 K, respectively.

We note that the comparison of our interferometric results to physical
models should be independent of the assumed distance (see Appendix
\ref{sec:dist}).
The inner radius is $\propto L^{1/2} \propto d$
in both the DDN and HSVK
models, and the linear sizes determined
from our interferometric results (converted from modeled angular sizes)
are also $\propto d$, and thus, the comparison is independent of $d$.


Flat accretion disk models \cite{HILLENBRAND+92}
are generally in good agreement
with the observed visibility data for early-type B-stars,
while puffed up inner disk models (DDN) seem more consistent
for later-type stars.  
We speculate that this could be due to different accretion mechanisms 
in earlier and later-type stars.  
A similar idea has been put forward based on the results of 
H$\alpha$ spectropolarimetry, where 
differences in the observations for early-type HBe stars and
later-type HAe stars have been attributed to a transition from disk
accretion in higher-mass stars to magnetic accretion in lower-mass stars
\citep{VINK+03}.

There is always the possibility that the visibilities for
some of the observed HAEBE sources may be (partially) due to
close companions.  For AB Aur and MWC 1080, we can rule out binary
models (with separations $\ga 1$ mas) with a high degree of confidence.
However, MWC 1080 is an eclipsing binary with a period of $P \approx 2.9$ days
\citep{SHEVCHENKO+94, CL99}.  The separation is much too small
to be detected by PTI, and the observed visibilities for this source
are thus probably due to an inclined circum-{\it binary} disk.
Observations over a time-span
of $\sim 100$ days (Table \ref{tab:obs}), with visibilities that are fairly
constant in time (Figures \ref{fig:v1685cyg-uv} and \ref{fig:v1685cyg-ha}) 
provide some evidence against V1685 Cyg being a binary.
As yet, the binarity status of
AS 442 and VV Ser remain uncertain based on our visibility data,
although radial velocity variations of spectral lines in AS 442
have been attributed to a binary with 
$P \approx 64$ days and $e \approx 0.2$ \citep{CL99}.

%

\section{Summary \label{sec:conc}}
We observed the HAEBE sources AB Aur, VV Ser, V1685 Cyg
(BD+40$^{\circ}$4124), AS 442, and MWC 1080 at 2.2 $\mu$m
with the Palomar Testbed Interferometer.
These are only the second published near-IR interferometric observations
of HAEBE stars.
From these high angular resolution data, 
we determined the angular size scales and orientations predicted
by uniform disk, Gaussian, ring, and accretion disk models,
and we examined whether the data were consistent with binary models.
AB Aur appears to be surrounded by a disk that is inclined by
$\la 30^{\circ}$, while VV Ser, V1685 Cyg, and MWC 1080 are
associated with more highly inclined circumstellar disks.
With the available data, we cannot distinguish between different
radial distributions, such as Gaussians, uniform disks, rings, or
accretion disks.

While the angular size scales determined in this work are generally
consistent with the only other near-IR interferometric measurements
of HAEBE stars by MST, our measurements
are the first that show evidence for significantly inclined 
morphologies.  Moreover, 
the derived angular sizes for 
early type Herbig Be stars in our sample, V1685 Cyg and MWC 1080, 
agree fairly well with those predicted by face-on accretion disk models
used by HSVK to explain observed spectral
energy distributions.
The observations of AB Aur, VV Ser, and
AS 442 are, however, not entirely compatible with these models,
and may be better explained
through the puffed-up inner disk models of DDN.

\noindent{ }

\noindent{\bf Acknowledgments.} The new data presented in this paper
were obtained at the Palomar Observatory using the Palomar Testbed
Interferometer, which is supported by NASA contracts to the Jet Propulsion
Laboratory.  Science operations with PTI are possible through the efforts 
of the PTI Collaboration 
({\tt http://huey.jpl.nasa.gov/palomar/ptimembers.html}) and Kevin Rykoski. 
This research made use of software produced by the Michelson Science Center
at the California Institute of Technology.
We thank R. Millan-Gabet for providing us with the IOTA
data, and for useful discussion.  We are also grateful to S. Metchev and
M. Konacki for obtaining the adaptive optics images and photometric data,
and to C. Koresko, B. Thompson, and G. van Belle for useful comments
on the manuscript.
J.A.E. and B.F.L. are supported by Michelson Graduate Research Fellowships.

\appendix
\section{Distance Estimates \label{sec:dist}}
AB Aur is associated with the Taurus-Auriga molecular cloud, and
thus the estimated distance to this source ($d = 140$ pc) is
accurate to $\sim 10\%$.
Photometric studies of VV Ser and other stars in Serpens
estimated distances of $d \approx 250$ pc \citep{CK+88} and 
$d \approx 310$ pc \citep{DL+91}, while an earlier study based on
photometry of a single source estimated $d = 440$ pc \citep{STROM+74}.  
Based on these estimates, we adopt a distance of 310 pc.
Distance estimates to V1685 Cyg range from 980 pc \citep[based on an
extinction-distance diagram for 132 stars within $3.5^{\circ}$;][]
{SHEVCHENCKO+91} to 1000 pc \citep[based on locating V1685 Cyg on the
main sequence;][]{STROM+72}, to 2200 pc \citep[based on photometry
of stars in a large scale region around V1685 Cyg;][]{HJ56}.  We
adopt a distance of $d=1000$ pc to V1685 Cyg, since the 2200 pc estimate
would imply a luminosity higher than expected for the published spectral
type.
AS 442 is associated with the
North American Nebula, and thus the adopted distance of 600 pc is
probably accurate to $\sim 10\%$.  The distance to MWC 1080 has been
determined by fitting photometric observations to the main sequence
\citep[$d = 1000$ pc;][]{HILLENBRAND+92}, and using the Galactic rotation curve
\citep[$d = 2500$ pc;][]{CANTO+84}.  
We adopt a distance of 1000 pc to MWC 1080,
since the 2500 pc estimate based on the Galactic rotation curve would imply
a luminosity much higher than expected for the published spectral type.
Moreover, the 2500 pc estimate is uncertain by $\sim 50\%$, while
the 1000 pc estimate is accurate to $\sim 20\%$.

\epsscale{0.7}
\begin{figure}
\plotone{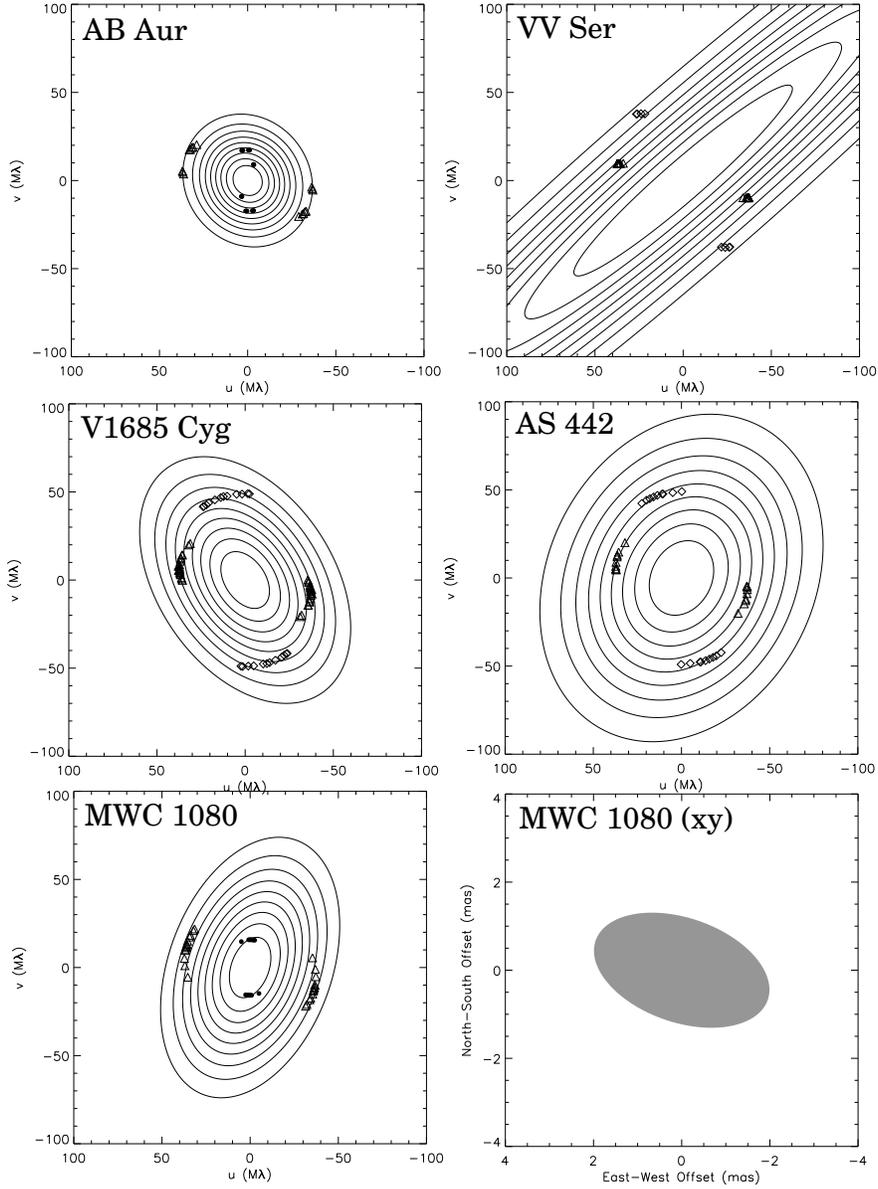}
\caption{Contour plots of best-fit inclined uniform disk models for AB Aur,
VV Ser, V1685 Cyg, AS 442, and MWC 1080, whose parameters
are listed in Tables \ref{tab:abaur}-\ref{tab:mwc1080}.
The contour increment is
10\% in $V^2$. We also plot the best-fit inclined disk model for MWC 1080
on the sky (bottom right panel, greyscale).
We over-plot the uv points sampled for each source by the PTI NW baseline
(open triangles), the PTI NS baseline (open diamonds), and by IOTA 
(filled dots).  
Since we know that the brightness
distributions of the sources are real, the visibilities must be reflection
symmetric (through the origin), and so we also plot these reflections of
the sampled uv points.
\label{fig:v1685cyg-model}}
\end{figure}

\clearpage
\begin{deluxetable}{lccccccccc}
\rotate
\tablewidth{0pt}
\tablecaption{Observed Sources \label{tab:sources}}
\tablehead{\colhead{Source} & \colhead{Alt. Name}  & \colhead{$\alpha$ (J2000)}
&  \colhead{$\delta$ (J2000)} &
\colhead{$d$ (pc)} & \colhead{Sp.Ty.} & \colhead{$V$}
& \colhead{$K$} & \colhead{$F^{\dag}_{\ast}$ (Jy)} &
\colhead{$F_{\rm x}^{\dag}$ (Jy)}}
\startdata
AB Aur & HD 31293 &  ${\rm 04^h55^m45.84^s}$ & 
${\rm +30^{\circ}33'04\rlap{.}''3}$ & 140 & A0pe & 7.07 & 4.27 &
1.92 & 10.59 \\
VV Ser & HBC 282 & ${\rm 18^h28^m49.00^s}$ & 
${\rm +00^{\circ}08'39\rlap{.}''0}$ & 310 & A0Vevp & 11.90 & 6.44 &
0.20 & 1.85 \\
V1685 Cyg & BD+40$^{\circ}$4124 & ${\rm 20^h20^m28.25^s}$ & 
${\rm +41^{\circ}21'51\rlap{.}''6}$ & 1000 & B2Ve & 10.71 & 5.70 &
0.42 & 3.64 \\
AS 442 & V1977 Cyg & ${\rm 20^h47^m37.47^s}$ & 
${\rm +43^{\circ}47'24\rlap{.}''9}$ & 600 & B8Ve & 10.89 & 6.75 &
0.20 & 1.21 \\
MWC 1080 & V628 Cas & ${\rm 23^h17^m26.10^s}$ & 
${\rm +60^{\circ}50'43\rlap{.}''0}$ & 1000 & B0eq & 11.68 & 4.83 & 0.87 & 9.85 
\enddata
\tablerefs{Distances, spectral types and V magnitudes from 
Hillenbrand et al. (1992), Mora et al. (2001), Strom et al. (1972),
de Lara et al. (1991), and Bigay \& Garner (1970). For discussion of
the adopted distances, see Appendix \ref{sec:dist}.
K magnitudes from the present work.
$\dag$: De-reddened fluxes.}
\end{deluxetable}

\clearpage
\begin{deluxetable}{lccccc}
\tablewidth{0pt}
\tablecaption{Summary of Observations \label{tab:obs}}
\tablehead{\colhead{Source} & \colhead{Date (MJD)} &
\colhead{Baseline}  & \colhead{ha cov.$^{\dag}$}
& \colhead{Calibrators (HD)}}
\startdata
AB Aur & 52575 & NW & [1.21,1.85] &  29645, 32301 \\
       & 52602 & NW & [-1.95,1.51] & 29645, 32301 \\
VV Ser & 52490 & NW & [-0.72,0.54] & 171834 \\
       & 52491 & NS & [-1.55,-0.74] & 171834 \\
       & 52493 & NW & [-1.31,0.20] & 171834 \\
       & 52499 & NW & [-0.96,0.84] & 164259,171834 \\
V1685 Cyg & 52418 & NW & [-1.10,-1.00] & 192640,192985 \\
          & 52475 & NW & [-1.69,1.44] & 192640,192985 \\
          & 52476 & NW & [-1.80,-0.48] & 192640,192985 \\
          & 52490 & NW & [-0.97,1.70] & 192640,192985 \\
          & 52491 & NS & [-1.27,2.38] & 192640 \\
          & 52492 & NW & [-0.90,-0.90] & 192640 \\
          & 52545 & NS & [-1.12,2.48] & 192640,192985 \\
AS 442 & 52475 & NW & [0.21,1.33] & 192640,192985 \\
       & 52476 & NW & [-0.21,-0.21] & 192640,192985 \\
       & 52490 & NW & [-1.11,0.38] & 192640 \\
       & 52491 & NS & [-0.69,2.54] & 192640 \\
       & 52492 & NW & [-1.05,0.00] & 192640 \\
       & 52545 & NS & [-0.87,1.58] & 192640,192985 \\
MWC 1080 & 52475 & NW & [0.17,0.17] & 219623 \\
         & 52476 & NW & [-1.99,0.52] & 219623 \\
         & 52490 & NW & [-0.14,1.39] & 219623 \\
\enddata
\tablerefs{${\dag}$: Hour angle coverage of the observations.}
\end{deluxetable}

\clearpage
\begin{deluxetable}{lccccccc}
\tablewidth{0pt}
\tablecaption{Properties of Calibrator Sources \label{tab:cals}}
\tablehead{\colhead{Name} & \colhead{$\alpha$ (J2000)}
&  \colhead{$\delta$ (J2000)} & \colhead{Sp.Ty.} & \colhead{$V$}
& \colhead{$K$} & \colhead{Cal. Size (mas)} & \colhead{$\Delta \alpha$
($^{\circ}$)}}
\startdata
HD 29645 & ${\rm 04^h41^m50.26^s}$ & ${\rm +38^{\circ}16'48\rlap{.}''7}$ &
G0V & 6.0 & 4.6 & $0.56 \pm 0.09$ & 8.2 \\
HD 32301 & ${\rm 05^h03^m05.75^s}$ & ${\rm +21^{\circ}35'23\rlap{.}''9}$ &
A7V & 4.6 & 4.1 & $0.47 \pm 0.10$ & 9.1 \\
HD 164259 & ${\rm 18^h00^m29.01^s}$ & ${\rm -03^{\circ}41'25\rlap{.}''0}$ &
F2IV & 4.6 & 3.7 & $0.77 \pm 0.08$ & 7.5 \\
HD 171834 & ${\rm 18^h36^m39.08^s}$ & ${\rm +06^{\circ}40'18\rlap{.}''5}$ &
F3V & 5.4 & 4.5 & $0.54 \pm 0.07$ & 6.8 \\
HD 192640 & ${\rm 20^h14^m32.03^s}$ & ${\rm +36^{\circ}48'22\rlap{.}''7}$ &
A2V & 4.9 & 4.9 & $0.46 \pm 0.02$ & 4.7$^1$,9.4$^2$ \\ 
HD 192985 & ${\rm 20^h16^m00.62^s}$ & ${\rm +45^{\circ}34'46\rlap{.}''3}$ &
F5V & 5.9 & 4.8 & $0.44 \pm 0.04$ & 4.3$^1$,5.9$^2$ \\
HD 219623 & ${\rm 23^h16^m42.30^s}$ & ${\rm +53^{\circ}12'48\rlap{.}''5}$ &
F7V & 5.6 & 4.3 & $0.54 \pm 0.03$ & 9.5
\enddata
\tablerefs{1,2: Offsets from V1685 Cyg, AS 442, respectively.}
\end{deluxetable}

\clearpage
\begin{deluxetable}{lcccc}
\tablewidth{0pt}
\tablecaption{Results of Modeling for AB Aur \label{tab:abaur}}
\tablehead{\colhead{Model} & \colhead{$\chi^2_{\rm r}$} &
\colhead{$\theta$ (mas)} & 
\colhead{$\psi$ ($^{\circ}$)} & \colhead{$\phi$ ($^{\circ}$)$^{\dag}$}}
\startdata
Face-On Gaussian & 2.42 & $3.59_{-0.08}^{+0.09}$ & & \\
Face-On Uniform & 1.78 & $5.34_{-0.09}^{+0.09}$ & & \\
Face-On Accretion & 1.93 & $2.18_{-0.02}^{+0.03}$ & & \\
Face-On Ring & 1.92 & $3.26_{-0.03}^{+0.02}$ & & \\
Inclined Gaussian & 0.96 & $3.88_{-0.27}^{+0.38}$ & $103_{-25}^{+23}$ & $35_{-18}^{+12}$ \\
Inclined Uniform & 0.89 & $5.80_{-0.45}^{+0.65}$ & $128_{-45}^{+30}$ & $26_{-19}^{+10}$ \\
Inclined Accretion & 0.88 & $2.30_{-0.11}^{+0.23}$ & $105_{-20}^{+34}$ & $27_{-17}^{+13}$ \\
Inclined Ring & 0.88 & $3.66_{-0.38}^{+0.42}$ & $144_{-51}^{+17}$ & $28_{-18}^{+10}$ \\
Binary Model & 8.96 & $3.41_{-0.28}^{+0.13}$ & $38_{-3}^{+7}$ & $0.58_{-0.03}^{+0.04}$ \\
\enddata
\tablerefs{$\dag$: For the binary model, $\phi$ represents the
brightness ratio, $R = F_2/F_1$.}
\end{deluxetable}

\clearpage
\begin{deluxetable}{lcccc}
\tablewidth{0pt}
\tablecaption{Results of Modeling for VV Ser \label{tab:vvser}}
\tablehead{\colhead{Model} & \colhead{$\chi^2_{\rm r}$} &
\colhead{$\theta$ (mas)} & 
\colhead{$\psi$ ($^{\circ}$)} & \colhead{$\phi$ ($^{\circ}$)$^{\dag}$}}
\startdata
Face-On Gaussian & 9.13 & $2.33_{-0.09}^{+0.09}$ & & \\
Face-On Uniform & 6.91 & $3.68_{-0.12}^{+0.12}$ & & \\
Face-On Accretion & 8.33 & $1.49_{-0.05}^{+0.05}$ & & \\
Face-On Ring & 5.86 & $2.30_{-0.07}^{+0.07}$ & & \\
Inclined Gaussian & 0.85 & $2.56_{-0.13}^{+1.66}$ & $37_{-55}^{+6}$ & $89_{-50}^{+1}$ \\
Inclined Uniform & 0.85 & $3.94_{-0.17}^{+2.35}$ & $41_{-53}^{+2}$ & $82_{-43}^{+8}$ \\
Inclined Accretion & 0.85 & $1.62_{-0.98}^{+1.58}$ & $38_{-70}^{+5}$ & $83_{-45}^{+7}$ \\
Inclined Ring & 0.85 & $2.44_{-0.11}^{+1.92}$ & $43_{-78}^{+5}$ & $81_{-51}^{+9}$ \\
Binary Model & 0.85 & $8.80_{-0.95}^{+1.02}$ & $176_{-3}^{+9}$ & $0.45_{-0.04}^{+0.38}$ \\
\enddata
\tablerefs{$\dag$: For the binary model, $\phi$ represents the
brightness ratio, $R = F_2/F_1$.}
\end{deluxetable}

\clearpage
\begin{deluxetable}{lcccc}
\tablewidth{0pt}
\tablecaption{Results of Modeling for V1685 Cyg \label{tab:v1685cyg}}
\tablehead{\colhead{Model} & \colhead{$\chi^2_{\rm r}$} &
\colhead{$\theta$ (mas)} & 
\colhead{$\psi$ ($^{\circ}$)} & \colhead{$\phi$ ($^{\circ}$)$^{\dag}$}}
\startdata
Face-On Gaussian & 6.51 & $1.96_{-0.11}^{+0.11}$ & & \\
Face-On Uniform & 7.52 & $3.17_{-0.15}^{+0.16}$ & & \\
Face-On Accretion & 6.75 & $1.27_{-0.06}^{+0.06}$ & & \\
Face-On Ring & 8.10 & $1.92_{-0.09}^{+0.09}$ & & \\
Inclined Gaussian & 2.32 & $2.43_{-0.37}^{+0.44}$ & $125_{-28}^{+9}$ & $51_{-16}^{+12}$ \\
Inclined Uniform & 2.36 & $3.91_{-0.55}^{+0.60}$ & $124_{-24}^{+9}$ & $50_{-14}^{+11}$ \\
Inclined Accretion & 2.33 & $1.57_{-0.22}^{+0.27}$ & $124_{-22}^{+9}$ & $50_{-14}^{+12}$ \\
Inclined Ring & 2.38 & $2.33_{-0.29}^{+0.37}$ & $122_{-24}^{+10}$ & $49_{-13}^{+11}$ \\
Binary Model & 3.33 & $3.41_{-0.56}^{+0.37}$ & $62_{-11}^{+6}$ & $0.24_{-0.04}^{+0.08}$ \\
\enddata
\tablerefs{$\dag$: For the binary model, $\phi$ represents the
brightness ratio, $R = F_2/F_1$.}
\end{deluxetable}

\clearpage
\begin{deluxetable}{lcccc}
\tablewidth{0pt}
\tablecaption{Results of Modeling for AS 442 \label{tab:as442}}
\tablehead{\colhead{Model} & \colhead{$\chi^2_{\rm r}$} &
\colhead{$\theta$ (mas)} & 
\colhead{$\psi$ ($^{\circ}$)} & \colhead{$\phi$ ($^{\circ}$)$^{\dag}$}}
\startdata
Face-On Gaussian & 0.99 & $1.49_{-0.19}^{+0.19}$ & & \\
Face-On Uniform & 1.04 & $2.44_{-0.28}^{+0.29}$ & & \\
Face-On Accretion & 0.99 & $0.95_{-0.12}^{+0.13}$ & & \\
Face-On Ring & 1.07 & $1.55_{-0.17}^{+0.17}$ & & \\
Inclined Gaussian & 0.94 & $1.63_{-0.29}^{+0.82}$ & $60_{-60}^{+120}$ & $41_{-41}^{+49}$ \\
Inclined Uniform & 0.94 & $2.67_{-0.34}^{+1.29}$ & $63_{-63}^{+117}$ & $39_{-39}^{+51}$ \\
Inclined Accretion & 0.94 & $1.03_{-0.18}^{+0.57}$ & $63_{-63}^{+117}$ & $36_{-36}^{+54}$ \\
Inclined Ring & 0.95 & $1.70_{-0.28}^{+0.80}$ & $65_{-65}^{+115}$ & $38_{-38}^{+52}$ \\
Binary Model & 0.95 & $2.69_{-1.50}^{+0.69}$ & $30_{-19}^{+32}$ & $0.21_{-0.01}^{+0.79}$ \\
\enddata
\tablerefs{$\dag$: For the binary model, $\phi$ represents the
brightness ratio, $R = F_2/F_1$.}
\end{deluxetable}

\clearpage
\begin{deluxetable}{lcccc}
\tablewidth{0pt}
\tablecaption{Results of Modeling for MWC 1080 \label{tab:mwc1080}}
\tablehead{\colhead{Model} & \colhead{$\chi^2_{\rm r}$} &
\colhead{$\theta$ (mas)} & 
\colhead{$\psi$ ($^{\circ}$)} & \colhead{$\phi$ ($^{\circ}$)$^{\dag}$}}
\startdata
Face-On Gaussian & 56.33 & $2.34_{-0.05}^{+0.05}$ & & \\
Face-On Uniform & 42.04 & $3.84_{-0.07}^{+0.07}$ & & \\
Face-On Accretion & 54.24 & $1.54_{-0.03}^{+0.03}$ & & \\
Face-On Ring & 36.00 & $2.33_{-0.05}^{+0.04}$ & & \\
Inclined Gaussian & 3.21 & $2.61_{-0.08}^{+0.11}$ & $71_{-9}^{+11}$ & $56_{-5}^{+6}$ \\
Inclined Uniform & 2.54 & $4.13_{-0.10}^{+0.12}$ & $70_{-8}^{+10}$ & $53_{-5}^{+7}$ \\
Inclined Accretion & 3.07 & $1.69_{-0.05}^{+0.07}$ & $71_{-9}^{+10}$ & $55_{-3}^{+5}$ \\
Inclined Ring & 2.28 & $2.47_{-0.06}^{+0.06}$ & $69_{-9}^{+10}$ & $51_{-6}^{+6}$ \\
Binary Model & 9.32 & $2.57_{-0.18}^{+0.22}$ & $56_{-3}^{+4}$ & $0.36_{-0.02}^{+0.02}$ \\
\enddata
\tablerefs{$\dag$: For the binary model, $\phi$ represents the
brightness ratio, $R = F_2/F_1$.}
\end{deluxetable}

\clearpage
\begin{deluxetable}{lcccc}
\tablewidth{0pt}
\tablecaption{Comparison with Hillenbrand et al. (2001) Models 
\label{tab:acc}}
\tablehead{\colhead{Source} & \colhead{$R_{\rm face-on}^{\ast}$} & 
\colhead{$R_{\rm inclined}^{\ast}$} &  
\colhead{$R_{\dot{M}=0}$} & \colhead{$R_{\dot{M} \ne 0}$} \\
& (AU) & (AU) & (AU) & (AU)}
\startdata
AB Aur & $0.15 \pm 0.01$ & $0.16 \pm 0.01$ & 0.07 & 0.12 \\ 
VV Ser & $0.23 \pm 0.01$ & $0.25 \pm 0.19$ & 0.03 & 0.13 \\
V1685 Cyg & $0.64 \pm 0.03$ & $0.79 \pm 0.13$ & 0.44 & 0.71 \\
AS 442 & $0.29 \pm 0.04$ & $0.31 \pm 0.12$ & 0.10 &   \\
MWC 1080 & $0.77 \pm 0.02$ & $0.85 \pm 0.03$ & 0.79 & 0.79 \\
\enddata
\tablerefs{$\ast$: Error bars based on 1-$\sigma$ uncertainties
of best-fit face-on and inclined accretion disk models.}
\end{deluxetable}

\clearpage
\begin{deluxetable}{lcccc}
\tablewidth{0pt}
\tablecaption{Comparison with Dullemond et al. (2001) Models 
\label{tab:ring}}
\tablehead{\colhead{Source} & \colhead{$R_{\rm face-on}^{\ast}$} &
\colhead{$R_{\rm inclined}^{\ast}$} & 
\colhead{$R_{2000}$} & \colhead{$R_{1500}$} \\
& (AU) & (AU) & (AU) & (AU)}
\startdata
AB Aur & $0.23 \pm 0.01$ & $0.26 \pm 0.03$ & 0.32 & 0.57 \\ 
VV Ser & $0.36 \pm 0.01$ & $0.38 \pm 0.14$ & 0.24 & 0.42 \\
V1685 Cyg & $0.96 \pm 0.05$ & $1.17 \pm 0.16$ & 3.71 & 6.59 \\
AS 442 & $0.47 \pm 0.05$ & $0.56 \pm 0.13$ & 0.52 & 0.93 \\
MWC 1080 & $1.17 \pm 0.02$ & $1.24 \pm 0.03$ & 8.69 & 15.45 \\
\enddata
\tablerefs{$\ast$: Error bars based on 1-$\sigma$ uncertainties
of best-fit face-on and inclined ring models.}
\end{deluxetable}


\begin{thebibliography}{99}
\aasup{Bigay \& Garnier 1970}{BG70}{Bigay, J.H., \& Garnier, R. 1970}{1}{15B}
\spie{Boden et al. 1998}{BODEN+98}{Boden, A.F., Colavita, M.M.,
van Belle, G.T., \& Shao, M. 1998}{3350}{872}
\bibitem[Born \& Wolf 1999]{BW99}{Born, M., \& Wolf, E. 1999,
Principles of Optics, 7 (Cambridge, UK:Cambridge University Press, 1999)}
\apj{Cant\'{o} et al. 1984}{CANTO+84}{Cant\'{o}, J., Rodr\'{i}guez, L.F.,
Calvet, N., \& Levreault, R.M. 1984}{282}{631}
\aa{Chavarr\'{i}a-K. et al. 1988}{CK+88}{Chavarr\'{i}a-K., C., de Lara, 
E., Finkenzeller, U., Mendoza, E.E., \& Ocegueda, J. 1988}{197}{151}
\apj{Chiang \& Goldreich 1997}{CG97}{Chiang, E.I., \& Goldreich, P. 1997}
{490}{368}
\apj{Colavita et al. 1999}{COLAVITA+99}{Colavita, M.M., et al. 1999}{510}{505}
\pasp{Colavita 1999}{COLAVITA99}{Colavita, M.M. 1999}{111}{111}
\aa{Corcoran \& Ray 1997}{CR97}{Corcoran, M., \& Ray, T.P. 1997}{321}{189}
\aasup{Corporon \& Lagrange 1999}{CL99}{Corporon, P., \& Lagrange, A.-M. 1999}
{136}{429}
\bibitem[Corporon 1998]{CORPORON98}{Corporon, P. 1998}Ph.D. thesis, Univ.
Joseph Fourier de Grenoble
\aa{de Lara et al. 1991}{DL+91}{de Lara, E., Chavarr\'{i}a-K., C., \&
L\'{o}pez-Molina, G. 1991}{243}{139}
\apj{DDN}{DDN01}{Dullemond, C.P.,
Dominik, C., \& Natta, A. 2001}{560}{957}
\aa{Eiroa et al. 2001}{EIROA+02}{Eiroa, C., et al. 2001}{384}{1038}
\apj{Grady et al. 1999}{GRADY+99}{Grady, C.A., Woodgate, B., Bruhweiler,
F.C., Boggess, A., Plait, P., Lindler, D.L., Clampin, M., \& Kalas, P. 1999}
{523}{L151}
\apj{Hartmann et al. 1993}{HARTMANN+93}{Hartmann, L., Kenyon, S.J., \&
Calvet, N. 1993}{407}{219}
\apjsup{Herbig 1960}{HERBIG60}{Herbig, G.H. 1960}{4}{337}
\apj{Herbst \& Shevchenko 1999}{HS99}{Herbst, W., \& Shevchenko, V.S. 1999}
{118}{1043}
\apj{HSVK}{HILLENBRAND+92}{Hillenbrand, L.A., Strom, S.E.,
Vrba, F.J., \& Keene, J. 1992}{397}{613}
\apjsup{Hiltner \& Johnson 1956}{HJ56}{Hiltner, W.A., \& Johnson, H.L. 1956}
{2}{389}
\mnras{Lynden-Bell \& Pringle 1974}{LBP74}{Lynden-Bell, D., \& Pringle,
J.E. 1974}{168}{603}
\apj{Mannings \& Sargent 2000}{MS00}{Mannings, V., \& Sargent, A.I. 2000}
{529}{391}
\apj{Mannings \& Sargent 1997}{MS97}{Mannings, V., \& Sargent, A.I. 1997}
{490}{792}
\nature{Mannings et al. 1997}{MKS97}{Mannings, V., Koerner, D.W., \&
Sargent, A.I. 1997}{388}{555}
\mnras{Mannings 1994}{MANNINGS94}{Mannings, V. 1994}{271}{587}
\aa{Meeus et al. 2001}{MEEUS+01}{Meeus, G., Waters, L.B.F.M., Bouwman, J.,
van den Ancker, M.E., Waelkens, C., \& Malfait, K. 2001}{365}{476}
\apj{MST}{M-G+01}{Millan-Gabet, R., Schloerb, F.P.,
\& Traub, W.A. 2001}{546}{358}
\apj{Millan-Gabet et al. 1999}{M-G+99}{Millan-Gabet, R., Schloerb, F.P.,
Traub, W.A., Malbet, F., Berger, J.P., \& Bregman, J.D. 1999}{513}{L131}
\apj{Miroshnichenko et al. 1999}{MIROS+99}{Miroshnichenko, A., Ivezic, Z.,
Vinkovic, D., \& Elitzur, M. 1999}{520}{L115}
\aa{Mora et al. 2001}{MORA+01}{Mora, A., et al. 2001}{378}{116} 
\aa{Natta et al. 2001}{NATTA+01}{Natta, A., Prusti, T., Neri, R., Wooden, D.,
Grinin, V.P., \& Mannings, V. 2001}{371}{186}
\bibitem[Natta et al. 2000]{NGM00}Natta, A., Grinin, V.P., \&
Mannings, V. 2000, Protostars and Planets IV, 559
\apj{Natta et al. 1993}{NATTA+93}{Natta, A., Palla, F., Butner, H.M.,
Evans, N.J., II, \& Harvey, P.M. 1993}{406}{674}
\aa{Oudmaijer et al. 2001}{OUDMAIJER+01}{Oudmaijer, R.D., et al. 2001}
{379}{564}
\aa{Pi\'{e}tu et al. 2003}{PDK03}{Pi\'{e}tu, V., Dutrey, A., \& Kahane, C.
2003}{in press}{}
\bibitem[Shevchencko et al. 1994]{SHEVCHENKO+94}Shevchencko, V.S., 
Grankin, N., Ibragimov, M.B., Kondratiev, V.Y., \& Melnikov, S. 1994,
in ``The nature and evolutionary status of Herbig Ae/Be stars, eds.
P.S. Th\'{e}, M.R. P\'{e}rez, \& E.P.J. van den Heuvel, 62, 43
\bibitem[Shevchencko et al. 1992]{SHEVCHENCKO+91}Shevchencko, V.S.,
Ibragimov, M.A., \& Chernysheva, T.L. 1991, SvA, 35, 229 
\aj{Skrutskie et al. 1996}{SKRUTSKIE+96}{Skrutskie, M.F., Meyer, M.R.,
Whalen, D., \& Hamilton, C. 1996}{112}{2168}
\apss{Steenman \& Th\'{e} 1991}{ST91}{Steenman, H., \& Th\'{e}, P.S. 1991}
{184}{9}
\apj{Strom et al. 1974}{STROM+74}{Strom, S.E., Grasdalen, G.L., \& 
Strom, K.M. 1974}{191}{111}
\apj{Strom et al. 1972}{STROM+72}{Strom, K.M., Strom, S.E., Breger, M.,
Brooke, A.L., Yost, J., Grasdalen, G.L., \& Carrasco, L. 1972}{173}{L65}
\mnras{Vink et al. 2003}{VINK+03}{Vink, J.S., Drew, J.E., Harries, T.J.,
\& Oudmaijer, R.D. 2003}{in press}{}
\end{thebibliography}
\end{document}